\documentclass[preprint,eqsecnum,aps]{revtex4}
\usepackage{graphicx,here}
\usepackage{amssymb, amsmath}
\usepackage{fancybox}

\font\bb=msbm10
\def \R {\hbox{\bb R}}

\newcommand{\width}{12cm}
\newcommand{\seqn}{\begin{equation}}
\newcommand{\eeqn}{\end{equation}}
\newcommand{\seqna}{\addtocounter{equation}{1}\begin{eqnarray}}
\newcommand{\eeqna}{\end{eqnarray}}

\newcommand{\Eq}[1]{Eq.\ (\ref{eq:#1})}
\newcommand{\Eqs}[2]{Eqs. (\ref{eq:#1}) and (\ref{eq:#2})}

\newcommand{\Sec}[1]{Sect.\ \ref{sec:#1}}

\newcommand{\lr}[1]{\left ( #1 \right )}
\newcommand{\lrb}[1]{\left [ #1 \right ]}

\newcommand{\prob}[2]{ \mbox{P} (#1 | #2)}
\newcommand{\pdf}[2]{ f (#1 | #2)}
\newcommand{\prior}[1]{ \mbox{P} (#1)}
\newcommand{\priorpdf}[1]{ \pi (#1)}

\newcommand{\model}[1]{\mbox{M}}

\newcommand{\loss}[1]{{\cal L}(#1)}
\newcommand{\risk}{{\cal R}}
\newcommand{\erisk}{{\cal R}_{\mbox{emp}}}

\newcommand{\EA}[1]{<#1>}
\newcommand{\ave}[1]{\EA{#1}}
\newcommand{\Ave}[2]{\EA{#1}_{#2}}

\newcommand{\var}[1]{\mbox{Var}(#1)}

\newcommand{\cov}[1]{\mbox{Cov}(#1)}

\newcommand{\cin}[1]{\lrb{l(#1), u(#1)}}

\newcommand{\Prob}[1]{\Pr[#1]}

\newcommand{\set}[2]{\{#1_1,\cdots,#1_{#2} \}}
\newcommand{\bX}{\mathbf X}
\newcommand{\bx}{\mathbf x}
\newcommand{\bz}{\mathbf z}
\newcommand{\btheta}{\bf \theta}
\newcommand{\htheta}{\hat{\theta}}
\newcommand{\bomega}{\bf \omega}

\newcommand{\D}[1]{D_{#1}}

\newcommand{\Chisq}[2]{\mbox{Chisq}(#1, #2)}
\newcommand{\Poisson}[2]{\mbox{Poisson}(#1, #2)}
\newcommand{\Binomial}[3]{\mbox{Binomial}(#1, #2, #3)}
\newcommand{\Gaussian}[3]{\mbox{Gaussian}(#1, #2, #3)}
{\obeylines

\gdef\looker{\ifx\next\lineending\vskip-1sp%
\noindent\else%
\ifx\next\end\vskip-1sp\let^^M\ \else%
\vskip0pt\noindent\fi\fi}

\gdef\lineending{\parfillskip=0pt plus1fil\relax\futurelet\next\looker}

\gdef\algorithm{\medskip
\def\note##1{\parfillskip=0pt\hfill##1}\def\ {\quad}%
\baselineskip=12pt\parskip=2pt\obeylines\let^^M=\lineending}

\gdef\endalgorithm{\medskip}

\gdef\algorithm{\bgroup\obeylines\obeyspaces\def\ {\quad}
\def\note##1{\parfillskip=0pt\hfill##1}
\leftskip=2pc\relax\let^^M=\lineending\relax}

\gdef\endalgorithm{\medskip\egroup}

}

\begin{document}

\title{Probability and Statistical Inference
\footnote{{\em SERC School in Particle Physics}, Chandigarh, India,
7-27 March, 2005}}

\author{Harrison B. Prosper}

\affiliation{Department of Physics, Florida State University,
Tallahassee, Florida 32306, USA}

\date{\today}

\begin{abstract}
These lectures introduce key concepts in probability and
statistical inference at a level suitable for graduate students 
in particle physics. 
Our goal is to paint as vivid a picture as possible of the concepts covered. 
\end{abstract}

\maketitle

\tableofcontents

\newpage
\section{Lecture 1 - Probability Theory, Part I}
\label{sec:Introduction}
Sir Harold Jeffreys~\cite{Jeffreys} begins his book, {\em Theory of
Probability}, with these words
\begin{quote}
    ``The fundamental problem of scientific progress, and a fundamental
one of everyday life, is that of learning from experience.''
\end{quote}
In everyday life, we learn from experience in a way that is still deeply
mysterious. However, in scientific research the learning
is more readily formalized:
we collect data in a systematic way about some aspect of the world
and, from these data, infer something of interest using more or less formal
methods. Probability theory is useful at all stages.

Given its central role in statistical
inference, we believe it is helpful to have a clear 
understanding of what probability
is and how that notion arose. Accordingly, these lectures are divided
into two parts: Lectures 1 and 2 cover probability theory, while Lectures
3 and 4 deal with statistical inference. In the first lecture, 
we begin with a sketch of the history of probability. This is followed
by a discussion of the elements of deductive and inductive reasoning,
ending 
with a discussion of some elementary aspects of probability theory.

\subsection{Historical Note}
The theory of probability arose from the ancient and popular 
pastime of gambling. One of the earliest references to chance, and
to the woes of
gambling, occurs in the cautionary
tale of {\em King Nala} from the epic poem {\em Mahabharata}~\cite{History}.
King Nala lost his kingdom in a gambling contest and was reduced to
working for King Bhangasuri as a chariot-driver. One day, while on a journey
with the king, Nala boasted of his mastery of horses. The king did not
take too kindly to such boasting and reminded Nala 
that no man knows everything. To make his point, the king made a quick 
estimate of the number of fruit on a nearby tree, the extraordinary
accuracy of which was verified by Nala, who counted the fruit one by
one. Nala
pleaded with the king to divulge the method that yielded such an
astonishingly accurate estimate. The king replied: 
\begin{quote}
``Know that I am
a knower of the secret of the dice and therefore adept in the art of
enumeration.''
\end{quote}
In the end, the king relented and told Nala the secret. 
It would seem from this tale that 
some notions of chance were understood, at least by some, 
in ancient times. However, probability theory as a recognizable 
mathematical discipline was established only centuries later.

In 1654, the French nobleman, Chevalier de M\'{e}r\'{e}, complained to
Blaise Pascal that the rules of arithmetic must be faulty.
His reason: the observation that his two methods of 
placing bets, using dice, 
did not work equally well, contrary to his expectation. He would bet on the
basis of 
obtaining at least one 6 in 4 throws of a single die, or, 
at least one double 6 in 24 throws of two dice. 
Pascal worked out the probabilities and showed that
the first outcome was indeed slightly more probable than the second. Thus
was born the mathematical theory of probability. 
 
By the late 17th century, probability was interpreted in several ways:
\begin{itemize}
\item as the fraction of favorable outcomes in a set
of outcomes considered {\em equally likely},
\item as a measure of {\em uncertain knowledge} of outcomes,
\item as a physical {\em tendency} in things that exhibit chance.
\end{itemize}
James Bernoulli (1654--1705) labored hard to make sense of these
different aspects of probability, but, dissastisfied with his labors, 
he chose not to publish his results.
Happily, however, in 1713, his nephew Nicholas Bernoulli
published {\em Ars Conjectandi} ({\em The Art of Conjecture}), 
James Bernoulli's famous treatise
on probability. This book contains the proof of 
an important result, namely, 
the {\bf weak law of large
numbers}, which we discuss later in this lecture.
Some decades later, the  English cleric  
Thomas Bayes (1702--1761) read (via a proxy!) the following paper before
the Royal Society, on 23 December, 1763: {\em An Essay towards solving a
Problem in the Doctrine of Chances}. This paper is notable for
at least two reasons. Firstly, in it, a proof is given of a
special case of what
became known as Bayes' theorem. Secondly, this paper makes explicit use
of probability as a measure of uncertain knowledge about something, 
in this case, uncertain knowledge of the value of a probability!
The ideas of Bayes, and probability theory, in general, 
were brought to great heights by Pierre Simon de Laplace (1749--1827) in
his book of 1812 entitled: 
{\em Th\'{e}orie Analytique des Probabiliti\'{e}s}. In it, amongst
other things, one finds the
general form of Bayes' theorem. One also finds results that soon became
controversial; indeed, that became the object of ridicule. Laplace made
extensive use of Bayes' theorem, sometimes in ways that yielded odd results.
From one of his results (the {\em law of succession}) one would conclude that
a 9-year old boy has a lesser chance of reaching the age of 10 than does
a 99-year old man to reach the age of 100. 
The logician George Boole was particularly 
scornful of Laplace's use of Bayes'
theorem.
In the Bayes-Laplace view of probability, the foundation
of the {\bf Bayesian approach} to statistical inference, probability
is construed 
as a measure of
the {\em plausibility} of an assertion. For
example, Bayes and Laplace would have had no difficulty 
with the assertion
``There is a 60\% chance of rain tomorrow''. 

For Boole and other
mathematicians and philosophers, however, 
the notion of probability as a measure of 
uncertain knowledge, or
the plausibility, of the truth of an assertion
seemed metaphysical and therefore unscientific. They 
therefore sought a
different interpretational foundation for the theory of probability,
grounded, as they perceived it, more firmly in experience.
As a result of the critiques of the Bayes-Laplace methods, and the
growing ``ideology of the objective'' in the natural sciences~\cite{Daston}, 
these 
methods fell into disfavor. This was not only because of 
discomfort with the
inherent subjectivity of the probabilities manipulated by Bayes and Laplace,
but because of the seemingly arbitrary
manner in which they assigned certain probabilities.
To excise such alleged
defects in the theory of probability a  
different approach was developed, which,
at the start of the 20th century, became the foundation of what
 has come to be known as the
{\bf frequentist approach} to statistical inference.
The newer approach, which comprises the body of
statistical ideas with  which most physicists are
familiar, is closely associated with the names of
Sir  Ronald   Aylmer  
Fisher (1890--1962), Jerzy Neyman, Pearson, Cramer,  
Rao, Mahalanobis, von Mises and Kolmogorov, to name but a 
few~\cite{Fisher}. The frequentist approach is
typically presented as if it were a single coherent 
school of thought. In fact, however, within this approach
views differed, sometimes sharply. Indeed, 
the sharpest disagreements were between Jerzy Neyman 
and Ronald Fisher, the two principal architects of the 
frequentist approach.
 
Fisher and Neyman, along with the other frequentists, 
did, however,  agree on one crucial point:
probability is to interpreted not as a measure of plausibility, or
uncertain knowledge, or
{\bf degree of belief}, but 
rather as the {\bf  relative frequency} with which something
happens, or will happen~\cite{RelativeFrequency}.
From the frequentist point of view, 
statements such as ``There is a  60\% chance of rain tomorrow'' 
are devoid of empirical content. Why? Because it is not possible 
 to repeat the day that is tomorrow and
count how often it rained.  By contrast, 
the statement ``There is a 60\% chance
of rain on days named March 7th'' is judged meaningful because
such days repeat and we can, therefore, assess by enumeration  
the relative frequency with
which it rains on days so named.

The frequentist viewpoint took hold in the physical sciences and became
the norm in particle physics~\cite{Cousins}. 
Indeed, that viewpoint is so entrenched
in our field that, until fairly recently, it was
hardly recognized that one has
a choice about how to conduct statistical inferences. However,  
during the latter half
of the 20th century the methods of Bayes and Laplace
have undergone a renaissance
initiated, in large measure, by  Sir Harold Jeffreys~\cite{Jeffreys}
(1891--1989) and vigorously developed by
like-minded physicists and
mathematicians,
notably Cox, de Finetti, Lindley, Savage and
Jaynes~\cite{Cox, DeFinetti, Box, Jaynes}. 
Moreover, after a somewhat slow start, beginning
with a few papers in the 1980s~\cite{Helene, Harrison, Harrison2}
a similar
renaissance is underway in particle physics~\cite{Durham}.

\subsection{Reasoning}
\label{sec:Reasoning}
\begin{quote}
    ``Probability theory is nothing but common sense reduced to calculation."

    --- Laplace, 1819
\end{quote}
Aristotle, who lived around 350 BC, 
was one of the first thinkers to attempt a formalization of
reasoning. He noticed that on those rare occasions when we reason correctly
we did so according to rules that can be reduced to the
syllogisms:
\vskip 24pt
\noindent
\begin{tabular}{l|c|c}
                & {\bf modus ponens} (ponere=affirm)
                & {\bf modus tollens} (tollere=deny)    \\ \hline
Major premise   &   If $A$ is TRUE, then $B$ is TRUE
                &   If $A$ is TRUE, then $B$ is TRUE    \\
Minor premise   &   $A$ is TRUE
                &   $B$ is FALSE \\
Conclusion      &   Therefore, $B$ is TRUE
                &   Therefore, $A$ is FALSE
\end{tabular}
\vskip 12pt

\noindent
In addition, if the statement $A$ is TRUE then its negation, 
written as $\overline{A}$, is, of necessity, 
FALSE. The statement $A$ is said to contradict $\overline{A}$.
A simple mnemonic for the syllogisms are the set of symbolic expressions:
\vskip 24pt
\noindent
\begin{tabular}{l|c|c}
                & {\bf modus ponens}    & {\bf modus tollens}   \\ \hline
Major premise   &   $AB = A$            &   $AB = A$            \\
Minor premise   &   $A  = 1$    &   $B  = 0$    \\
Conclusion      &   $B  = 1$    &   $A  = 0$    \\
\end{tabular}
\vskip 12pt

\noindent
The symbols $A$, $B$, $1$, $0$, and their
negations, $\overline{A}$, $\overline{B}$, $\overline{1}$ and
$\overline{0}$, are variously referred to as
events, statements, assertions, or {\bf propositions}. 
The symbol $1$ represents a proposition that is always TRUE; 
the symbol $0$, its negation, is 
always FALSE.

Here is a simple example.
Let $A = \mbox{\em She finished school}$ and let
$B = \mbox{\em She is educated}$. Our major premise is: 
If {\em She finished school} is TRUE then {\em She is educated} is TRUE. 
Suppose that our minor premise is
{\em She finished school} is TRUE. We may, as a matter of logic, 
conclude that {\em She is educated} is TRUE. 
On the other hand, however, if the proposition $B$ 
is TRUE, that is, {\em She is educated} it does {\em not} follow that
$A$ is TRUE, that is, that {\em She finished school}.
She may be educated because she is self-taught!
Conversely, if
$A$ is FALSE, that is, {\em She finished school} is FALSE, 
we cannot logically conclude that $B$ is FALSE, that is, {\em She is educated}
is FALSE.
But, if {\em She is educated} is, in fact, FALSE then we
can conclude that
{\em She finished school} is FALSE. 

These logical arguments can be readily constructed using the symbolic
expressions and  noting that if $B$ is set to $1$ (that is, to
the proposition that is always
TRUE) in $AB = A$ we get $A = A$ and we are no wiser about the
truth or falsity of $A$. Likewise, if $A = 0$, that is, $A$ is FALSE, 
then the truth or falsity of
$B$ cannot be ascertained. 

Deductive reasoning, as we have just sketched, is extremely powerful; witness
the immense scope and power of mathematics. However, to learn from experience
we need a way to reason as it were ``backwards'', that is, to reason
{\em inductively}.
In the example above, suppose it is true
that {\em She is educated}. We acknowledge the possibility that we
could be wrong, but, it is certainly {\em plausible} that if
{\em She is educated} is, in fact, true 
this renders the proposition {\em She finished school}
more likely. The methods of
Bayes and Laplace can be viewed as a formalization of this mode of
{\bf plausible reasoning}. Indeed, the 
Bayes-Laplace theory, and its subsequent developments by Sir Harold
Jeffreys, Cox, Jaynes and others, can be viewed as an extension of
logic to include truth
values that lie between FALSE and TRUE. Moreover, if one
makes the idealization that truth values can
be represented by real numbers in the interval $[0,1]$, it can
be shown that these
numbers satisfy
the axioms of probability and, as such, are a quantitative measure of the
plausibility of propositions. 
 These arguments assign a quantitative meaning to the weaker syllogisms:

\vskip 24pt
\noindent
\begin{tabular}{l|c|c}
Major premise   &   If $A$ is TRUE, then $B$ is TRUE
                &   If $A$ is TRUE, then $B$ is TRUE    \\
Minor premise   &   $B$ is TRUE
                &   $A$ is FALSE \\
Conclusion      &   Therefore, $A$ is more plausible
                &   Therefore, $B$ is less plausible.
\end{tabular}
\vskip 12pt

\subsection{Probability Calculus}
\label{sec:ProbabilityCalculus}
The theory of probability can be founded in many different ways. One way, 
is to regard probability as a {\em function} with range [0,1], defined
on sets of events or propositions. But in order speak of sets of
propositions, we need to know how they are to be manipulated;
that is, we need an algebra of propositions. The appropriate algebra, 
{\bf Boolean algebra}, 
was invented by George Boole (1854).
If $A$, $B$, $C$, $1$, $0$
and their negations are propositions, and $+$ and $\cdot$ are binary operations
then, one form of the axioms--the {\bf Huntington axioms}--is 

\vskip 24pt
\noindent
\begin{tabular}{l|rcl|rcl}
                    &   $A + 0$         & = &    $A$
                    &   $A \cdot 1$     & = &    $A$  \\
                    &   $A + \overline{A}$      & = &    $1$
                    &   $A \cdot \overline{A}$  & = &    $0$  \\
Commutativity law   &   $A \cdot B$                    & = &   $B \cdot A$
                    &   $A + B$                 & = &   $B + A$  \\
Distributivity law  &   $A \cdot (B + C)$       & = &   $A \cdot B + A \cdot C$
                    &   $A + B \cdot C$        & = &  $(A + B)\cdot (A + C)$ \\
\end{tabular}
\vskip 12pt

\noindent
Usually, we drop the ``$\cdot$'' operator in expressions to simplify
the notation. From these axioms the theorems of Boolean algebra
can be deduced as logical
consequences. 
\vskip 24pt
\framebox{{\bf Exercise}: Prove the theorems below.}
\vskip 12pt
\begin{tabular}{l|rcl|rcl}
                    &   $A + 1$         & = &    $1$
                    &   $A  0$     & = &    $0$  \\
                    &   $\overline{0}$  & = &    $1$
                    &   $\overline{1}$  & = &    $0$  \\
                    &   $A + AB$                & = & $A$
                    &   $A(A + B)$              & = & $A$ \\
Idempotency law     &   $A  A$                    & = &   $A$
                    &   $A + A$                 & = &   $A$    \\
Associativity law   &   $A (B C)$       & = & $(A B) C$
                    &   $A + (B + C)$           & = & $(A + B) + C$ \\
de Morgan's laws    & $\overline{A  B}$ & = & $\overline{A} + \overline{B}$
                & $\overline{A + B}$ & = & $\overline{A} \, \overline{B}$.
\end{tabular}
\vskip 12pt
\noindent
Consider the propositions $A$, $B$, $A+B$ and $AB$, to each of which
we (somehow) have assigned the numbers $P(A)$, $P(B)$, $P(A+B)$ and
$P(AB)$. The axioms of probability specify how these numbers are related.
Let $A$ and $B$ be the propositions $A = \mbox{\em It will rain today}$ and
$B = \mbox{\em It is the rainy season}$, respectively. The probability
of $A$ {\em given} $B$, written thus $P(A|B)$, that is, the
probability it will rain today {\em given} that it is the rainy season,
is defined by
\begin{equation}
P(A|B) \equiv \frac{P(AB)}{P(B)}.
\end{equation}
The number $P(A|B)$ is called the
{\bf conditional probability} of $A$ given $B$. Note that $P(B)$ is the
probability of $B$ {\em without} restriction, while $P(A|B)$ is the
probability of $A$ when we {\em restrict} to the circumstance
in which $B$ is true. Strictly speaking, there is a restriction on $B$ also;
$B$ is true given some other more encompassing 
circumstance $C$. Probabilities are always context-dependent numbers. 
There is no such thing
as the probability to create a $t\bar{t}$ pair; there is, however,
the probability to create a $t\bar{t}$ {\em given} some particular set of
conditions. Therefore, we should, 
in principle, always make the conditioning explicit and write every
probability
in the form $P(A|C)$. In practice, if the conditioning is clear we may
drop it from the notation.

The other set of probability axioms can be taken to be the
{\bf product rule}
\seqna
    \prob{AB}{C} & = & \prob{B}{AC} \prob{A}{C}, \nonumber \\
                 & = & \prob{A}{BC} \prob{B}{C},
\label{eq:prule}
\eeqna
and the
{\bf sum rule}
\seqn
    \prob{A}{C} + \prob{\overline{A}}{C} = 1,
\label{eq:srule}
\eeqn
and the conventions
\seqna
    \prob{1}{C} & = & 1, \nonumber \\
    \prob{0}{C} & = & 0.
\label{eq:crule}
\eeqna
As an illustration of the use of the rules given above we
prove a theorem that relates
$\prob{A+B}{C}$ to $\prob{A}{C}$ and $\prob{B}{C}$.
We need merely to apply the above rules repeatedly:
\seqna
    \prob{A + B}{C}
    & = &1 - \prob{\overline{A + B}}{C}       \nonumber \\
    & = &1 -  \prob{\overline{A} \, \overline{B}}{C}  \nonumber \\
    & = &1 - \prob{\overline{B}}{\overline{A}C} \prob{\overline{A}}{C}
\nonumber \\
    & = &1 - \lrb{1 - \prob{B}{\overline{A}C}} \prob{\overline{A}}{C}
\nonumber \\
    & = &1 - \prob{\overline{A}}{C} +
\prob{B}{\overline{A}C} \prob{\overline{A}}{C} \nonumber \\
    & = &\prob{A}{C} + \prob{B}{\overline{A}C} \prob{\overline{A}}{C}
\nonumber \\
    & = &\prob{A}{C} + \prob{\overline{A}B}{C} \nonumber \\
    & = &\prob{A}{C} + \prob{\overline{A}}{BC} \prob{B}{C} \nonumber \\
    & = &\prob{A}{C} + \lrb{1 - \prob{A}{BC}} \prob{B}{C} \nonumber \\
    & = &\prob{A}{C} + \prob{B}{C} - \prob{A}{BC} \prob{B}{C} \nonumber \\
    \prob{A+B}{C} & = &\prob{A}{C} + \prob{B}{C} - \prob{AB}{C}.
\eeqna
The Huntington axioms seem intuitively reasonable, but the
product and sum rules, Eqs.~(\ref{eq:prule}) and (\ref{eq:srule}),
seem less so. Remarkably, these rules can be derived from the more
primitive axioms:
\begin{itemize}
\item
{\bf Axiom 1)} Plausibilities $q$ can be represented by real numbers.
\item
{\bf Axiom 2)} The plausibilities $q(B)$ and $q(A|B)$ of
a proposition $B$ and that of another $A$ {\em given}
the first determine the plausibility $q(AB)$
of the joint proposition $AB$; that is, $q(AB)$
is some function of $q(B)$ and $q(A|B)$.
\item
{\bf Axiom 3)} The
plausibility $q(A)$ of a proposition $A$ determines
the plausibility $q(\overline{A})$ of its converse $\overline{A}$.
\end{itemize}
 This was first done by the physicist,
R.T.~Cox~\cite{Cox}, in 1946, who
showed 
that 
plausibilities or degrees of belief follow rules that are isomorphic
to those of probability and thus provide a {\bf subjective interpretation}
of the latter. Moreover, 
well before Cox's theorem,
James Bernoulli, who, along with his contemporaries, regarded the subjective 
interpretation 
of probability as self-evidently sensible~\cite{Daston}, proved a theorem
that provides a link between relative frequency 
and the abstraction we call probability.

\subsection{Objective Interpretation}
\label{sec:RelativeFrequency}
In the {\bf objective interpretation},
probability is interpreted as
the  relative frequency with which something happens, or
could happen. Let $n$ be
the number of experiments or {\bf trials}; for example, this could be
the number of proton-proton collisions at the LHC.
Let $k$ be the number of {\em successes}; for example, it could
be the count in a
given mass bin of Higgs boson events.
The relative frequency of successes is
\seqn
    \frac{k}{n} \, .
\eeqn
It is a matter of experience that
as $n$ grows ever larger
the relative frequency $k/n$ settles down to a number, call it $p$,
whose natural interpretion is
the probability of a success. Unfortunately, this interpretation
is not quite as straightforward as it seems.
Any theory of probability that {\em defines} the latter
as the limit of $k/n$ must contend with the following
possibility.
It is possible that on every trial
we get a success, or a failure, or we alternate between the
two {\em ad infinitum}. It is important, therefore, to be precise
about what is meant by the {\em limit} of the (rational) number $k/n$.
The correct statement, first noted
by James
Bernoulli (1703), is the 
{\bf weak law of large numbers}, mentioned briefly above. This theorem
states that
\seqn
    \lim_{n \rightarrow \infty} 
\Prob{|\frac{k}{n} - p| > \epsilon} = 0 \, ,
\eeqn
for any real number $\epsilon > 0$. That is,
 as the number of trials goes to infinity,
the {\em probability} $\Prob{*}$,
that the relative frequency
$k/n$ differs from the {\em probability}
$p$ by more than $\epsilon$,  becomes vanishingly small.

The implied recursion in this theorem is conceptually problematic. 
If, indeed, probability is to be defined as {\em nothing more} than
the limit of a relative frequency,
then the two
probabilities that occur in Bernoulli's theorem must both be
limits of
relative frequencies. 
 The second probability $p$ in the theorem may legitimately be
viewed as the ``limit'' of the relative frequency $k/n$. 
However, to define the first probability
$\Prob{*}$ requires a {\em second} application of Bernoulli's theorem.
But that second application 
will specify yet another $\Prob{*}$, which must itself be defined
in terms of a limit, and so it goes. It would seem that
we cannot avoid being ensnared in an infinite hierarchy of 
infinite sequences of trials. Moreover, never, in practice, do we
ever conduct infinite sequences of trials and therefore the limit, as
it true of all limits, is an abstraction.

\subsection{Subjective Interpretation}
\label{sec:Subjective}
We can avoid the infinite hierarchy of trials if
we are prepared to interprete
the first probability in Bernoulli's theorem 
differently from the second. If we
interpret the first as a measure of
plausibility then the theorem is a statement
about the plausibility of the proposition
$\lim_{n \rightarrow \infty} k/n = p$.
Bernoulli's theorem, as he himself interpreted it,
declares
 that it is
plausible to the point of certainty that $k/n \rightarrow p$ as
the number of trials grows without limit.
The import of this theorem, and Bernoulli's interpretation of it, is 
that probability as relative frequency
is a {\em derived} notion 
pertaining to a special class of circumstances, namely, those in
which one can entertain, in principle, 
performing {\em identically} repeated
trials in which the relative frequency converges, in the precise manner
of Bernoulli's theorem,
to some number $p$, which, because it satisfies the axioms
of probability, we are at liberty to call a probability. 
The Standard Model is an
example of a physical theory that can predict the limiting numbers $p$ for
the kind of identically repeated
trials performed in high energy physics experiments.

The position advocated here is that probability is an abstraction that can
be usefully interpreted in at least two different ways: as the
limit of a relative frequency
and as a degree of belief. Moreover, the first is best
understood in terms of the second.

\subsection{Bayes' Theorem}
\label{sec:BayesTheorem}
In 1763, Thomas Bayes published a paper
in which a
special case of a theorem, that bears his name, appeared.
Bayes' theorem
\seqn
    \prob{B_{k}}{AC} = \frac{\prob{A}{B_{k}C} \prob{B_{k}}{C}}
    {\sum_{i} \, \prob{A}{B_{i} C} \prob{B_{i}}{C}},
\eeqn
where $A$, $B_{k}$ and $C$ are propositions,
is a direct consequence of the
product rule, \Eq{prule}, of probability theory.
Consider
two propositions $A$ and $B$. They are said to be
{\em mutually exclusive} if the truth of one denies the
truth of the other, that is: $\prob{AB}{C} = 0$. In that case,
from the theorem we proved earlier, we conclude that
\seqn
    \prob{A+B}{C} = \prob{A}{C} + \prob{B}{C},
\eeqn
which is easily generalized to any number of mutually exclusive propositions.
A set of mutually exclusive
propositions $B_{k}$ is said to be {\em exhaustive} if their probabilities
sum to unity:
\seqn
    \sum_{k} \, \prob{B_{k}}{C} = 1.
\eeqn
Let $B_1$ and $B_2$ be exhaustive propopsitions. Consider the propositions
$AB_1$ and $AB_2$. From the product rule, we can write
\seqna
P(AB_1) & = & \prob{B_1}{A} P(A) , \\
P(AB_2) & = & \prob{B_2}{A} P(A) .
\eeqna
Now add the two equations
\seqna
P(AB_1) + P(AB_2) & = & \left[ \prob{B_1}{A} + \prob{B_2}{A} \right]
P(A), \\
& = & P(A).
\eeqna
This summation over exhaustive propositions is called
{\bf marginalization}, and is an extremely important operation in
probability calculations. 
If $B_{k} D_{j}$ are a set of mutually exclusive and exhaustive
joint propositions, then we can write
Bayes' theorem as
\seqn
    \prob{B_{k} D_{j}}{AC} = \frac{\prob{A}{B_{k} D_{j} C}
                             \prob{B_{k} D_{j}}{C}}
    {\sum_{i,l} \, \prob{A}{B_{i} D_{l} C} \prob{B_{i} D_{l}}{C}}.
\label{eq:bayest}
\eeqn
\vskip 24pt
\noindent
\framebox{{\bf Exercise}: Prove this form of Bayes' theorem.}
\vskip 24pt
Bayes' theorem is, of necessity,
true irrespective of how probabilities are
interpreted. Consider the following example.
A calorimeter shower arises either from an electron ($e$) or from a
jet ($j$). Some fraction of the energy of the incident object is
deposited in the electromagnetic calorimeter, often referred to
as the ``em-fraction''. We
impose the requirement $f \equiv \mbox{em-fraction} > 0.6$ and assume:
\vskip 24pt
\begin{tabular}{lccl}
$\prob{f}{e}$  & = & 0.90 & $\; \; \; \Prob{\mbox{electron to pass cut}}$,  \\
$\prob{f}{j}$  & = & 0.05 & $\; \; \; \Prob{\mbox{jet to pass cut}}$,       \\
$\prior{e}$    & = & 0.15 & $\; \; \; \Prob{\mbox{electron}}$,              \\
$\prior{j}$    & = & 0.85 & $\; \; \; \Prob{\mbox{jet}}$.
\end{tabular}
\vskip 24pt
\noindent
We wish to compute $\prob{e}{f}$, the probability that the shower was caused
by an electron, given that the em-fraction exceeds 0.6.
Applying Bayes' theorem we get
\seqna
    \prob{e}{f}  & = &
    \frac{\prob{f}{e} \prior{e}}
         {\prob{f}{e} \prior{e} + \prob{f}{j} \prior{j}} ,
                \nonumber \\
                                        & = &
    \frac{0.90 \times 0.15}{0.9 \times 0.15 + 0.05 \times 0.85},  \nonumber \\
                                        & = &   0.76.
\eeqna
We conclude that there is a 76\% probability that the shower is caused by
an electron. This calculation is correct whether or not the probabilities
are regarded as relative frequencies or degrees of belief.

\newpage
\section{Lecture 2 - Probability Theory, Part II}

\subsection{Probability Distributions}
\subsubsection{Random Variables}
Statisticians make a distinction between a {\bf random variable} $\bX$ and
its value $\bx$. A random variable can be thought of as 
a map $X$,
\seqn
        X:\Omega \rightarrow \R \, ,
\eeqn
between a set of possible {\bf events} or {\bf outcomes} 
$\Omega = \set{\omega}{N}$ 
and the set of reals $\R$. 
The map $X$ assigns a
real number $x = X(\omega)$, called the value of the random variable,  
to every outcome $\omega \in \Omega$.
The height of persons who pass you
in the street is an example of a random variable. Its
possible events are the people who can pass
you and its value is
the height of a person. Since the outcome is random so too
is the value of the random variable.
Note, however, that in spite of the name the
map $X$ itself is generally not random! Rather it is the set $\Omega$ of
possible outcomes that possesses the
(rather mysterious) quality called {\bf randomness}. One can think
of that property as a manifestation of a {\bf randomizing agent} whose job
it is to pick an outcome from the set of possibilities, according to a rule
that is not readily discernable.
The randomizing agent, however, need not be governed by chance!
Consider the set of possible outcomes $\Omega = \{0,\ldots,9\}$
and 
the
function $X$ that maps this set to the subset $\{0,\ldots,9\} 
\in \R$. Their exists a random variable $X$ whose
value is the
next decimal digit of $\pi$, starting, say, from 
the first. The digits of $\pi$ do not occur by chance even though they 
form an excellent random sequence.
The same is true of,
so-called, pseudo-random number generators, which provide sufficiently
random sequences of real numbers---indispensible in
Monte Carlo-based calculations, even though, again, 
the randomizing agent is not governed by chance; indeed, it is strictly
deterministic. 
Usually, a random variable is 
denoted by an upper case symbol, while one of its values is
 denoted by the corresponding lower
case symbol. Thus, if $\bX$ is a random variable then
$\bx$ denotes one of its values. 
However, for simplicity we shall not use this convention,
but refer to both with the same symbol.

\subsubsection{Properties}
In general, we are most interested in propositions involving real
numbers of the form 
$x \in (x_1, x_2)$. When $x$ is continuous, $P(x)$, is called
a {\bf probability distribution function}, while its derivative 
\seqn
\label{eq:pdf}
f(x) = \frac{dP(x)}{dx},
\eeqn
(assuming
it exists) is called a {\bf probability density function}. Notice that
probabilities, being pure numbers, are dimensionless, whereas densities have
dimensions $x^{-1}$. Note, also, that from the definition, \Eq{pdf},
\seqn
dP(x) = f(x) \, dx ,
\eeqn
and
\seqna
P(x) & = & \int dP(x), \\
& = & \int f(x) \, dx.
\eeqna

Given a probability distribution function $P(x)$, its {\bf moments}
$m_r(z)$ about a value $z$ is defined by
\seqna
m_r(z) & = & \int (x - z)^r dP(x) , \\
       & = & \int (x - z)^r f(x) \, dx .
\eeqna
Of particular importance are the first moment 
about zero and the second moment about
the first. The first moment about zero, $m_1(0)$, is called the {\bf mean}
and is often denoted by the symbol $\mu$. The second moment about the first,
that is about the mean, $m_2(\mu)$, is called the {\bf variance} of the
distribution. Its square-root, often denoted by the symbol $\sigma$, is
the {\bf standard deviation}, which is one measure of the width of the
distribution. The {\bf mode} of a probability density $f(x)$ is the value
of $x$ at which the density is a maximum. Finally, the {\bf median} of a
distribution is the value of $x$ that divides it into two equal parts. The
median is generally most meaningful if $x$ is a 1-dimensional variable. Note,
that if the density $f(x)$ is symmetrical about the mode, its mode, mean and
median coincide.

\subsubsection{Common Densities and Distributions}
Below we list the most commonly encountered 
densities and distributions, while in Fig.~1

\begin{tabbing}
----\=
------------------------------\=
-----------------------------------------\=
------------------\kill
  \> Uniform$(x, a, b)$   
  \> $1/(b-a)$                      
  \> $x \in [a,b]$ \\

  \> Binomial$(x, n, p)$  
  \> $\binom{n}{x} p^x (1-p)^{n-x}$ 
  \> $x \in [0,1,\cdots,n]$ \\

  \> Poisson$(x, a)$      
  \> $a^x \exp(-a) / x!$ 
  \> $x \in [0,1, \cdots)$ \\

  \> Gaussian$(x, \mu, \sigma)$      
  \> $\exp[-(x-\mu)^2 / 2\sigma^2] / \sigma \sqrt{2\pi}$ 
  \> $x \in (-\infty, +\infty)$ \\

  \> Chisq$(x, n)$      
  \> $x^{n/2 - 1}\exp(-x/2) / 2^{n/2}\Gamma(n/2)$ 
  \> $x \in [0, +\infty)$ \\

  \> Gamma$(x, a, b)$      
  \> $x^{b - 1} a^b \exp(- a x) / \Gamma(b)$ 
  \> $x \in [0, +\infty)$ \\

  \> Exp$(x, a)$      
  \> $a \exp(- a x)$
  \> $x \in [0, +\infty)$
\end{tabbing}
\noindent
we show examples of a few of them.
\begin{figure}[htbp]
\label{fig:pdf}
        \begin{center}
                \includegraphics[width=\width]{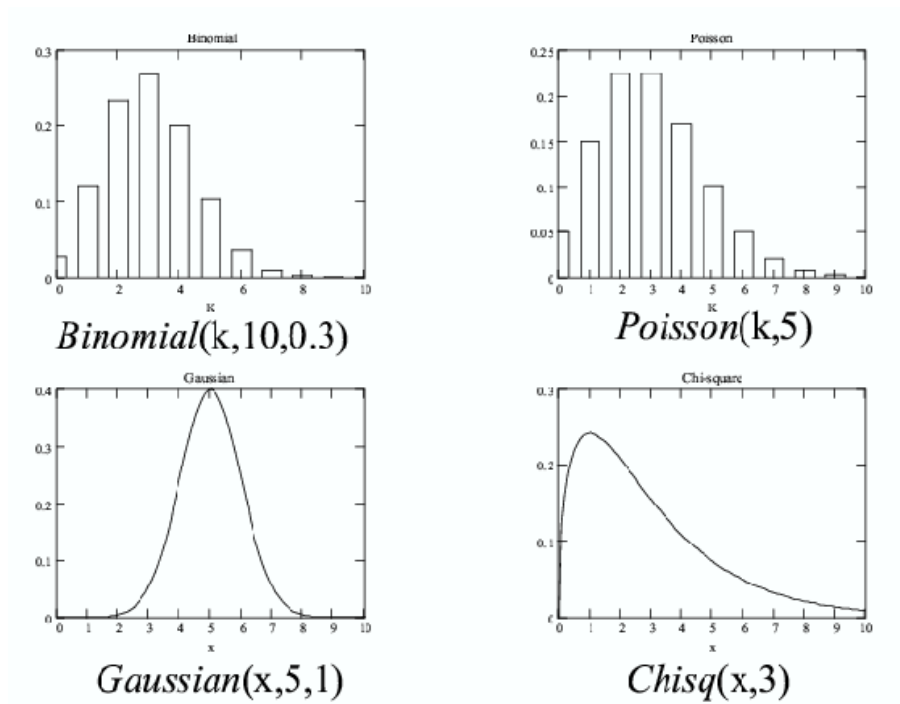}
                \caption{Examples of 
		  the most commonly used
		  distributions in particle physics.}
        \end{center}
\end{figure}

\noindent
\vskip 24pt
\framebox{
{\bf Exercise}: Calculate the mean and variance of each density, listed
above.
}

\subsection{The Binomial Distribution}
\label{TheBinomialDistribution} 
A {\bf Bernoulli} trial is one with only two outcomes, success ($S$) or
failure ($F$). Particle physicists conduct almost
perfect Bernoulli trials in which every collision, say 
between a proton and a proton
at the Large Hadron Collider, 
creates ($S$), or does not create ($F$), 
an event of interest. A 
success could be, for example, the creation of a Higgs boson event. 
Typically, we are interested in the probability $\prob{k}{n}$ of
$k$ successes given $n$ trials, or some function thereof.
Our task is to calculate this probability, from first principles.
Even if one is of the opinion that relative frequency is the only 
legitimate scientific way to think about probability, in practice it
is exceedingly difficult, if not impossible, 
to make headway, {\em from first principles}, using
this interpretation alone. Instead, 
we reproduce here an interesting result about
Bernoulli trials, due Bruno de
Finetti~\cite{DeFinetti}, following the presentation given by 
Heath and Sudderth\cite{HeathSudderth} and Caves~\cite{Caves}.

Suppose we have observed a 
sequence of Bernoulli trials $S_{k,n} = x_1,\ldots,x_n$, 
with $k$ successes in $n$ trials. We assume that these are the
{\em only} data of which we have knowledge.
We note that the 
probability we wish to calculate, $\prob{k}{n}$, makes no reference
to the particular sequence at hand. But,
to compute $\prob{k}{n}$, we must, nevertheless, be able to
assign a probability to a sequence of trials, a problem that, in general,
is extremely difficult.
However, given some crucial assumptions the
problem can be solved.

We assume that the details of the 
particular sequence observed are unimportant and that the 
only thing that matters is 
the total number of successes $k$ in the $n$ trials we have conducted.
We are therefore led to consider, not just the sequence we have
observed, but the set of all sequences of length $n$ with
$k$ successes, of which the one we observed is a particular instance. 
Denote by $P(S_{k,n,j})$ the probability of the $j^{\mathrm th}$
sequence $S_{k,n,j}$.
de Finetti~\cite{DeFinetti} argues that
the probabilities we assign, {\em at this 
stage}, must {\em of necessity} be
subjective. They are subjective in that they are based
on what we believe to be reasonable probability assignments, given
the objective information at hand, namely, 
the observed sequence of trials and their
outcomes. The probabilities we assign may be informed by predictions
from, say, the Standard Model or some theory beyond it, 
but we do not know {\em at this stage}
whether or not the predictions are correct.
After all, the trials are
being conducted precisely for the purpose of testing these predictions.

What then is the probability of $k$
successes in $n$ trials, regardless of the sequence? 
The answer, according to the rules of probability theory, is to 
add up all the probabilities $P(S_{k,n,j})$,
\begin{equation}
    P(k|n) = \sum_j P(S_{k,n,j}) ,
\end{equation}
that is, to  marginalize
over all the details that are deemed irrelevant; in this case, 
propositions of the form: the $j^{\mathrm th}$ sequence is $x_1,\ldots,x_n$.
Unfortunately, we can go no further  unless we are prepared to
introduce more assumptions. We shall make two more assumptions. The first is
that the order of
trials is irrelevant; more precisely, we assume that the
probability of a sequence of trials is symmetric with respect to all
permutations of the order of trials. Each sequence,
$S_{k,n,j}$, becomes, in effect, indistinguishable. Since
they are indistinguishable
we have no reason
to favor one sequence over another. In the absence of reasons to do otherwise
it would be rational to assign, to each sequence, the {\em same}
probability. Since there are $\binom{n}{k}$ indistinguishable
sequences, 
the probability of $k$ successes in $n$ trials, regardless of the sequence,
is
\begin{equation}
    P(k|n) = \binom{n}{k} P(S_{k,n}),
\end{equation}
where $S_{k,n}$ can be any one of the sequences $S_{k,n,j}$. The second
assumption is that the sequence $S_{k,n}$ can be embedded in
one or more {\em arbitrarily long} 
sequences $S_{r,m}$ of $r$ successes in $m \geq n$ trials in the
following way
\begin{equation}
\label{eq:ProbSequence}
    P(S_{k,n}) = \sum_{r=0}^{m} P(S_{k,n}|S_{r,m}) \ P(S_{r,m}).
\end{equation}
Sequences that satisfy both of these assumptions 
are said to be {\bf exchangeable}. The probabilities $P(S_{r,m})$
must still be freely assigned by us and, at present, there is
nothing more about them that can be said. However, the
exchangeability assumption yields a unique assessment of  $P(S_{k,n}|S_{r,m})$,
to which we not turn.

By assumption, all
successes are indistinguishable, as are all failures. Therefore, the
probability $P(S_{k,n}|S_{r,m})$ of $k$ successes and $n-k$ failures
in $n$ trials {\em given} that they are embedded in a
a sequence of $r$ successes and $m-r$ failures, in $m$ 
trials, is
akin to drawing, {\em without replacement}, $k$ red balls and $n-k$
white balls out of a box containing $r$ red balls plus $m-r$ white balls.
Since the sequences are indistinguishable, and that consequently
the order of trials is irrelevant, 
we can consider {\em any} convenient sequence to compute $P(S_{k,n}|S_{r,m})$,
such as the
one in which we get $k$ successes (red
balls) followed by $n-k$ failures (white balls). Noting that we
start with a box containing $m$ balls of which $r$ are red, the
probability to draw $k$ red balls is the product of $k$ fractions
\begin{equation}
    \left( \frac{r}{m}\right) \left(\frac{r-1}{m-1}\right)
    \cdots\left(\frac{r-(k-1)}{m-(k-1)}\right) =
    \frac{r!}{(r-k)!} \, / \, \frac{m!}{(m-k)!} \, ,
\end{equation}
while the probability to draw $n-k$ white balls from the remaining
$m-k$ balls of which $m-r$ are white is the product of $n-k$ fractions
\seqna
    \left(\frac{m-r}{m-k}\right) \left(\frac{m-r-1}{m-k-1}\right)
    \cdots \left(\frac{m-r-(n-k-1)}{m-k-(n-k-1)}\right) & = &
    \frac{(m-r)!}{(m-r-(n-k))!} \, \nonumber \\ 
    & / & \, \frac{(m-k)!}{(m-n)!} \, ,
\eeqna
which yields
\begin{equation}
\label{eq:ProbSeqSeq}
     P(S_{k,n}|S_{r,m}) = \frac{r!}{(r-k)!}
    \frac{(m-r)!}{(m-r-(n-k))!} \, / \, \frac{m!}{(m-n)!} \, .
\end{equation}
We can write \Eq{ProbSequence} as an integral
\begin{equation}
\label{eq:IntRep}
    P(S_{k,n}) = \int_0^1  P(S_{k,n}|S_{zm,m})  \, \pi_m(z) \, dz \, ,
\end{equation}
where
\begin{equation}
    \label{eq:ZDensity}
    \pi_m(z) \equiv \sum_{r=0}^m P(S_{zm,m}) \delta(z - r/m), 
\end{equation}
and $r/m$ is the observed relative frequency of success. By
assumption, we can make the sequences $S_{r,m}$ arbitrarily
long. When we do so, $P(S_{k,n}|S_{zm,m}) \rightarrow z^k
(1-z)^{n-k}$ as $m \rightarrow \infty$ and the functions $\pi_m(z)$ coalesce
into a continuous density $\pi(z)$. Putting together the pieces
we obtain de Finetti's Representation Theorem
\begin{equation}
\label{eq:RepTheorem}
    P(k|n) =  \int_0^1 \Binomial{k}{n}{z} \, \pi(z) \, dz,
\end{equation}
for Bernoulli trials. This remarkable result shows that for
exchangeable sequences of trials the probability $P(k|n)$ of $k$ successes
in $n$ trials is a binomial distribution weighted by a
density, $\pi(z)$. What exactly is $\pi(z)$? It is simply the probability
{\em we} have assigned to every sequence, characterized by the relative
frequency $z$. In other words, $\pi(z)$   
encodes {\em our} assessment of the likely
value of the relative frequency {\em in an infinite sequence
of trials}. If we knew, or we wished to act as if we knew, or we have
a prediction, that the relative frequency is $p$, then we would set
$\pi(z) = \delta(z-p)$, in which case
\Eq{RepTheorem} reduces to the binomial distribution. 

The important point to take away from this is that we have arrived at the
binomial distribution starting with {\em subjective} assessments of the 
probability of sequences of trials and the powerful assumption of
 exchangeability.

\subsection{The Poisson Distribution}
\label{ThePoissonDistribution} 
From the discussion above, it would seem that the binomial distribution is
the appropriate one to describe a typical high energy physics {\bf counting
experiment}. However, it is more usual to take note of the fact that 
the probability of a success $p << 1$. Given $n$ trials, the average number
of successes is $a = p n$. If we write
$\Binomial{k}{n}{p}$ in terms of $a = p n$ and take the limit 
$n \rightarrow \infty$, while keeping $a$ constant, it will tend towards 
$\Poisson{k}{a}$. Given that the probabilities $p$ are typically
very small, in practice
it is the Poisson distribution that is used to
describe the number of events observed or the count in a given bin of
a histogram.
\vskip 24pt
\framebox{{\bf Exercise}: 
Show that $\Binomial{k}{n}{p}$ $\rightarrow$ $\Poisson{k}{a}$
in the limit $p = a /n \rightarrow 0$.} 
\vskip 24pt
Another interesting way to understand the Poisson distribution is as
the outcome of a particular {\bf stochastic process}, which, roughly
speaking, is a system that evolves through {\em random} changes of
state. Suppose that at time $t+\Delta t$ we have recorded $k$ counts.
In a Poisson process one assumes 
that the probability to get a single count in the short
time interval $(t, t + \Delta t)$ is given by $q \Delta t$. Since this
probability is small, we can arrive at $k$ counts at time $t+\Delta t$
in {\em at most} two ways:
\begin{enumerate}
\item we had $k$ counts at time $t$ and recorded none in $(t, t + \Delta t)$,
\item we had $k-1$ counts at time $t$ and recorded 1 count in
$(t, t + \Delta t)$.
\end{enumerate}
Let 
\seqna
P_k(t+\Delta t) & = & \mbox{be the probability that the count is $k$ 
  at time $t+\Delta t$}, \\
P_k(t) & = & \mbox{be the probability that the count is $k$ 
  at time $t$}, \\
P_{k-1}(t) & = & \mbox{be the probability that the count is $k-1$ 
  at time $t$}, \\
q \Delta t & = & \mbox{be the probability of recording a {\em single} count in 
$(t, t + \Delta t)$}.
\eeqna
Given the two possible state changes from time $t$ to time $t+\Delta t$ we
deduce that the probabilities are related by the {\bf finite difference
equation}
\seqn
P_k(t+\Delta t) = (1 - q \Delta t) \, P_k(t) + q \Delta t \, P_{k-1}(t),
\eeqn
which can be re-expressed as 
\seqn
\frac{P_k(t+\Delta t) - P_k(t)}{\Delta t} = 
  - q P_k(t) + q P_{k-1}(t).
\eeqn
In the limit $\Delta t \rightarrow 0$, we obtain the differential
equation
\seqn
\label{eq:BirthDeath}
\frac{dP_k(t)}{d t} = 
  - q P_k(t) + q P_{k-1}(t),
\eeqn
which is a simple example of a 
{\bf birth - death equation}. (See Ref.~\cite{sampling} for another
example involving Poisson processes.) The first
term on the right-hand side describes the ``death'' rate, while the
second term describes the ``birth'' rate. Such equations describe the
probability of a given ``population'' size at time $t$.
\vskip 24pt
\framebox{
{\bf Exercise}: Solve \Eq{BirthDeath} and show that
$P_k(t)  = \Poisson{k}{qt}$, for $q = \mbox{constant}$.}
\vskip 24pt
\framebox{{\bf Exercise}: 
Repeat the calculation with $q(t) = \exp(-t/\tau) / \tau$.}

\subsection{The Gaussian Distribution}
\label{TheGaussianDistribution} 
The Gaussian distribution, also known as the {\bf normal distribution},
is the most important distribution in applied probability, principally
because of the {\bf Central Limit Theorem}, which roughly states that
\begin{quote}
{\em All reasonable distributions become Gaussian in the limit of
large numbers.}
\end{quote}
This is true, in particular, for the Poisson distribution. This is 
a result of practical
importance in that it is the basis of  $\chi^2$ methods to fit
functions to histograms and in the associated goodness-of-fit
tests (see Lecture 4). To illustrate this theorem,
first write
$\Poisson{k}{a}$ as $\exp[\ln \Poisson{a+x}{a}]$, in which we have set
$k = a + x$, and then allow $k \rightarrow \infty$. By using the approximation
\seqna
\ln \Poisson{k}{a} 
& = & k \ln a - a - \ln k! , \nonumber \\
& \approx & k \ln a - a - k \ln k + k - \ln \sqrt{2\pi k} , \\
\eeqna
one can show that the Poisson distribution 
becomes Gaussian when the counts become large.
\vskip 24pt
\framebox{{\bf Exercise:} Show that $\Poisson{k}{a} \rightarrow 
\Gaussian{k}{a}{\sqrt{a}}$.
}

\subsection{The $\chi^2$ Distribution}
\label{TheChiSqDistribution} 
The $\chi^2$ distribution is closely related to the Gaussian. Indeed, if 
$x_i \sim \Gaussian{x_i}{\mu_i}{\sigma_i}$, where $\mu_i$ and
$\sigma_i$ are known constants, then the quantity 
$z = \sum_{i=1}^n (x_i - \mu_i)^2 / \sigma_i^2$ has a $\chi^2$ density with
$n$ degrees of freedom~\cite{AccordingTo}.
An instructive way to
compute the density of $z$ is to use the intuitively clear 
formula~\cite{Gillespie}
\seqn
f(z) = \int \delta(z - h(x)) dP(x),
\eeqn
where $h(*)$ is some function of $x$, for example, 
$h(x) = \sum_{i=1}^n (x_i - \mu_i)^2 / \sigma_i^2$. 
The formula states that the density
$f(z)$ is given by the sum of the probabilities
$dP(x) = \prod_{i=1}^n f(x_i) dx_i$ over all values of $x_i$ consistent
with the constraint $z = h(x)$. By using the
integral representation of the $\delta$-function, 
\seqn
\delta(x) = \frac{1}{2\pi} \int_{-\infty}^{\infty} e^{i \omega x} d \omega,
\eeqn
we can write $f(z)$ as the Fourier
integral
\seqn
\label{eq:fprime}
f^{\prime}(z) = \frac{1}{2\pi i}\int_{-\infty}^{\infty} e^{i \omega z} 
F(\omega) d \omega ,
\eeqn
of the complex function
\seqn
\label{eq:Fomega}
F(\omega) = i \int e^{-i \omega h(x)} dP(x). 
\eeqn
If the exponential function in \Eq{Fomega} can be factorized into
a product of terms, each depending on a single variable $x_i$, it may be
possible to calculate $F(\omega)$ explicitly. This happens to be the
case for the function 
$h(x) = \sum_{i=1}^n (x_i - \mu_i)^2 / \sigma_i^2$. For this case, we
can write
\seqn
\label{eq:Fomega2}
F(\omega) = i \int dx_1 \, \Gaussian{x_1}{\mu_1}{\sigma_1} 
\cdots \int dx_n \, \Gaussian{x_n}{\mu_n}{\sigma_n} \, e^{-i \omega h(x)} , 
\eeqn 
which factorizes into a product of $n$ 1-dimensional integrals,
each of the same form. Using
the result $\int_{\infty}^{\infty} \exp[-(x-\mu)^2/ 2 \sigma^2] = \sigma 
\sqrt{2\pi}$, one finds
\seqn
F(\omega) = \frac{i}{(1+2 i \omega)^{n/2}}, 
\eeqn
which, from \Eq{fprime}, yields 
$z \sim \Chisq{z}{n}$.
\vskip 24pt
\framebox{{\bf Exercise}: Give a complete derivation of this result.
Hint: use contour integration.}
\vskip 24pt
For a more complex example of such a calculation, see Ref.~\cite{lifetime}.

\subsubsection{A Brief Word on Fitting}
The quadratic form $Q = \sum_{i=1}^n (x_i - \mu_i)^2 / \sigma_i^2$ is
commonly used to fit a function $\mu(\theta_1,\cdots,\theta_P)$, with
$P$ parameters $\theta_k, \, \, k = 1,\cdots,P$, to a histogram of
$n$ bins, with count $k_i$ in bin $i$. {\em If} the counts are
large enough (say $k > 10$), and {\em if} the variances $\sigma_i^2$ are
accurately known, then $Q \sim \Chisq{Q}{n-P}$ approximately. However, 
even if either, or both, conditions are not met $Q$ can still be used
to perform a fit, but its density will not be $\chi^2$, in general. Its
actual density, however, can be estimated by Monte Carlo simulation. The
density of $Q$ is typically used to test goodness-of-fit (see Lecture 4).

\newpage
\section{Lecture 3 - Statistical Inference, Part I}
\subsection{Descriptive Statistics}
One of the very first tasks in the analysis of data is to characterize
the data using a few numerical summaries. A {\bf statistic} is any
function of the data sample $\bx = x_1,\cdots x_n$. They can be as simple as
the {\bf sample average},
\seqn
\label{eq:SampleAverage}
\bar{x} = \frac{1}{n} \sum_{i=1}^n x_i,
\eeqn
and the {\bf mean squared error} (MSE),
\seqn
\label{eq:MSE}
\mbox{MSE} = \frac{1}{n} \sum_{i=1}^n (x_i - \bar{x})^2,
\eeqn
or as complex as the output of a full-blown analysis program. These summaries
provide a useful compression of the data, making it easier to gain some
understanding of the main features. 

\subsection{Ensemble Averaging}
In principle, before any serious analysis is undertaken a thorough exploration
of the behaviour of the proposed analysis method should be conducted. This
forms part of the {\bf experimental design} phase of an experiment. Such 
studies usually appear in Technical Design Reports (TDR).
The goal, in principle, is to ascertain, {\em a priori}, which analysis
method is best, in some agreed upon manner, with the intention of
applying the best method to the data when they are available.
In practice, however, such studies are done before, during, and after 
analyses of data. And often one decides, after the fact, which of
several analyses merit seeing the light of day.
Whatever the motivation, and stage of the analysis, there is broad agreement
that it is
crucial to
study the behaviour of methods on an {\bf ensemble} of 
artificial data samples, usually created by Monte Carlo simulation. These
studies are often referred to as {\bf ensemble tests}. As a simple
illustration, we discuss the ensemble behaviour of a few
simple statistics.
 
In general, each sample $\bx = x_1,\cdots,x_n$
within the ensemble will yield a different value
for the average, \Eq{SampleAverage}. Intuitively, we expect these
averages to be closer to the mean of the distribution, from which the data
have been generated, than the individual data $x_1,\cdots, x_n$ that comprise
each average. Given some measure of ``closeness'' to
the mean it would be natural to compute its average value 
over the ensemble; that is, to perform an 
{\bf ensemble average}, denoted by the symbol $<\cdots>$, of the
closeness measure.
Consider first the ensemble average of the sample average, \Eq{SampleAverage},
\seqna
\EA{\bar{x}} 
& = & \EA{ \frac{1}{n} \sum_{i=1}^n x_i }, \nonumber \\
& = & \frac{1}{n} \sum_{i=1}^n \EA{ x_i }, \nonumber \\
& = & \frac{1}{n} n \mu, \nonumber \\
& = & \mu .
\eeqna
We have assumed that the $x_i$ are identically distributed, in which case
$\EA{ x_i } = \mu$, and that the {\bf bias}, 
\seqn
\label{eq:Bias}
b \equiv \EA{ x } - \mu, 
\eeqn
is zero. Take as our measure of closeness to the mean $\mu$ the
square of 
\seqn
\Delta \bar{x} = \frac{1}{n} \sum_{i=1}^n \Delta x_i,
\eeqn
where the {\bf error}, $\Delta x_i = x_i - \mu$. 
Squaring both sides, and taking the
ensemble average, yields
\seqn
\EA{ \Delta \bar{x}^2 } = \frac{1}{n^2} \sum_{i=1}^n 
\sum_{j=1}^n \cov{x_i,x_j}, 
\eeqn
where $\cov{x_i, x_j} \equiv \EA { \Delta x_i \Delta x_j }$ is called
the {\bf covariance matrix}. If this matrix is diagonal, the data are
said to be {\bf uncorrelated}. However, this does {\em not} necessarily
imply that they are independent; that is, that the probability 
distribution $P(x)$, generating the samples, is of the form
$dP(x) = \prod_{i=1}^n f(x_i) \, dx_i$. If the $x_i$ are independent in
this sense then they are of necessity uncorrelated, but the converse is
not true; uncorrelated data may, or may not, be independent. The diagonal
elements $\var{x_i} \equiv \EA{ \Delta x_i^2 }$, which can be
written as $\var{x_i} = \EA{x_i^2} - \EA{x_i}^2$, are the {\bf variances}.
Note that the MSE, \Eq{MSE}, the bias and the variance are
related as follows
\seqn
\mbox{MSE} = b^2 + \var{x}.
\eeqn

The common practice is to use ensembles whose samples are
independent and therefore uncorrelated. 
However, for practical reasons it may be necessary to
use an ensemble in which the correlation between samples is not quite zero.
This will be the case in an ensemble in which the samples are
generated by a {\bf bootstrap}
method~\cite{Barlow}. In a bootstrap method one draws many samples of size $n$ 
from a population $x_1,\cdots,x_m$ of size $m \geq n$. Each sample
is created by drawing elements
$x_i$, one at a time --- at random and {\em with replacement}, from the
finite population. Since the samples are drawn with replacement, they will
in general have elements $x_i$ that are common. 
Consequently, any statistic calculated from them will be 
correlated across the ensemble. In particular, the sample averages will
be correlated. In the following we shall assume this to be the case.

We can re-write $\EA{ \Delta \bar{x}^2}$ as follows
\seqna
\EA{ \Delta \bar{x}^2 } 
& = & \frac{1}{n^2} \sum_{i=1}^n \sum_{j=1}^n \EA { \Delta x_i \Delta x_j },
\nonumber \\
& = & \frac{1}{n^2} \sum_{i=1}^n \EA { \Delta x_i^2 }  
  + \frac{1}{n^2}  \sum_{i=1}^n \sum_{j \neq i}^n 
\EA { \Delta x_i \Delta x_j },
\nonumber \\
\label{eq:CrossTerms}
& = & \frac{\sigma^2}{n}  
  + \frac{1}{n^2}  \sum_{i=1}^n \sum_{j \neq i}^n 
\EA { \Delta x_i \Delta x_j },
\eeqna
assuming zero bias and variance $\sigma^2 = \EA{ \Delta x_i^2}$. If the
samples are uncorrelated then the cross-terms in \Eq{CrossTerms}
average to zero and we obtain the
well-known result that the variance of the average, $\bar{x}$, 
is smaller by a factor $n$ than the variance of $x$, confirming that
the average is indeed closer to the mean $\mu$ than is $x$. Suppose, however,
that the cross-terms do not vanish and each is given by 
$\EA { \Delta x_i \Delta x_j } = \rho \sigma^2$, where $\rho \in (-1,+1)$ 
is the {\bf correlation coefficient}. For this simple case we find
\seqn
\label{eq:CorAve}
\EA{ \Delta \bar{x}^2 }  =  \frac{\sigma^2}{n}\lrb{1 + (n-1) \rho}.
\eeqn
As expected, correlated samples yield less precise averages.
And, unlike averages from uncorrelated samples, increasing the sample size $n$ 
indefinitely does not help  since according to \Eq{CorAve} the variance 
of the average has a lower bound of $\rho \sigma$.
 
\subsection{Estimators}
As noted in Lecture 1, our goal as scientists 
is to learn from experience by conducting carefully controled experiments
that yield data from which we can {\em infer} something interesting about
the system under investigation.
Given a {\bf data-set}  $\bx = \set{x}{N}$,  
a mathematical model  $M$, characterized by the parameters
$\btheta$, and the associated probability
$\prob{\bx}{\btheta}$ we  
use statistical inference to decide the best values to
assign   to   the    
parameters     $\btheta$.  If we   have   several   models
$M_1, M_2 \ldots$  then we may, in
addition, wish to decide which  one is best. This, of course,
presupposes
that we know what we  mean by {\em best}.
The mapping
$\set{x}{N} \longrightarrow \set{\theta}{M}$ from our data-set
to the parameters, or to the set of models, 
is an example of a {\bf decision function}, which will be
denoted by the symbol
$d$. Suppose that our model depends upon a single
parameter $\theta$. Denote by $\htheta$ any {\bf estimate} thereof.
If the decision function is such that 
$\htheta = d(\bx)$ then the function $d$
is called an {\bf estimator}
for $\theta$. 
One can think of the estimator as a program, which
when data are entered into it outputs estimates. 
The estimator
could be as simple as an averaging operation 
or as complex as several full-scale analysis program.

\subsection{Loss and Risk}
\label{sec:LossAndRisk}

To choose a decision function we need a way to 
quantify the quality of the associated
decisions. In general, every decision, especially bad ones, entail some
loss. The loss can be quantified with
a {\bf loss function}, $\loss{\theta,d}$, which
depends on both 
the decision function and the parameter being estimated. 
The idea of a loss function is useful in both frequentist and Bayesian
analysis. However, the two approaches use the loss function
differently:
\begin{itemize}
\item {\bf Frequentist:} In making {\em inferences}
data we could have  
observed are as relevant as data observed.
\item {\bf Bayesian:} In making {\em inferences}, 
only the data observed are relevant.
\end{itemize}
Accordingly, in the frequentist 
approach we consider the loss pertaining to every data-set which could
have been observed,
as well as the loss pertaining to the data actually
obtained. In the Bayesian theory, on the other hand, 
all possible hypotheses must be considered in light of the data-set
actually obtained.

In either case,  the desire to average the loss function in some way
motivates the definition of a new function 
\seqn
    \risk = \Ave{\loss{\theta,d}}{*}, \label{eq:riskdef}
\eeqn
called the {\bf risk function}, where the subscript $*$ denotes averaging with 
respect to either $\bx$ or $\theta$. 
In one case, the averaging is done 
with respect to 
all possible data-sets $\bx$ for fixed $\theta$ (frequentist), while
in the other the averaging is done
with respect to all possible $\theta$ for fixed $\bx$ (Bayesian). 
In the frequentist approach, the risk
function is an ordinary function of the parameter $\theta$ but
a functional of the decision function $d$; that is, it
depends on the set of all possible values of $d$. 
In the Bayesian approach, the risk function is
a functional of $\theta$. However, it is generally not regarded as 
a function of $\bx$ because the data are considered to be constants.

It should not be construed from the above that Bayesians do not care about
data-sets that could have been observed. On the contrary, it is 
absolutely essential during the {\em design} of an experiment, or of 
an analysis, 
to consider what could be observed 
in order to conduct the best possible experiment or the most effective
analysis. In the Bayesian approach, however,
when the time comes to make inferences only the data
actually acquired are deemed relevant.

\subsection{Risk Minimization}
\label{sec:RiskMinimization}
A statistical analysis can be viewed as a procedure that
minimizes a risk function in order to arrive at an optimal
decision, usually an optimal decision about the value of a parameter
or a model. In particle physics, one often speaks
of ``optimizing an analysis''. What we are doing, without being explicit
about it, is minimizing {\em some} unstated risk function.
If the risk function is known then, in principle, an optimal
decision can be had with respect to the underlying loss function.
However, in many circumstances although
the loss function is known, since we choose it, the risk function is not.
In these cases, we must make do with an estimate of the risk function, the
most common of which is given by
\seqn
        \erisk = \frac{1}{n}\sum_{i=1}^n \loss{\theta,f(\bx_i, \bomega)},
\eeqn
where $f(\bx_i, \bomega)$ is a suitably parameterized function, with
parameters $\bomega$ and data $\bx_i$,
that one hopes is flexible enough to include a good approximation to the
optimal decision function $d$, say at the point
$\bomega = \bomega_0$. The function $\erisk$ is called
the {\bf empirical risk function}. Its minimization, to obtain
an approximation to the optimal decision function $d$, is
a widely used strategy in data analysis, encompassing
everything from curve-fitting to the training of sophisticated 
learning machines.. The strategy is referred to as
{\bf empirical risk minimization}.

The most important mathematical property of
empirical risk, and the property that makes it useful in practice
is that the function $f(\bx_i, \bomega_0)$, found by minimizing
the empirical risk, is expected
to converge to the optimal decision function  $d(\bx)$
as the sample size $n$ goes to infinity, provided that the function
 $f(\bx, \omega)$ is sufficiently flexible and the minimization
algorithm is effective at finding the minimum.

\subsection{The Bayesian Approach}
\label{sec:BayesianTheory}
The Bayesian approach to statistical inference 
is firmly grounded in the subjective interpretation
of probability. 
 Whereas the frequentist approach deals only with the distributional
properties of data, that is, with statements
of the form
\seqn
        \prob{Data}{Theory} \, ,
\label{eq:DataTheory}
\eeqn
the Bayesian approach admits, in addition, 
statements of the form
\seqn
        \prob{Theory}{Data} \, ,
\label{eq:TheoryData}
\eeqn
that is, 
 the probability that a given {\em Theory}
is true, in light of evidence provided by {\em Data}.
This is precisely the kind of statement that most physicists would wish to
make.
The connection between the two probabilities, \Eqs{DataTheory}{TheoryData}, 
is given by Bayes' theorem, \Eq{bayest},
\seqn
\prob{Theory}{Data} =
\prob{Data}{Theory}
\, \prior{Theory} / \prior{Data} .
\label{eq:bayestheorem}
\eeqn
The probability $\prior{Theory}$ is called the
{\bf prior probability}. 
It encodes what we believe we know about the {\em Theory} 
independently of the {\em Data}. 
The probability  $\prob{Data}{Theory}$ is
sometimes referred to, loosely, as the {\bf likelihood}, while the probability
$\prob{Theory}{Data}$ is called the {\bf posterior probability}. More
correctly, the likelihood is a function $\propto \prob{Data}{Theory}$.
Viewed this way, it is not a probability.
  
The power of the Bayesian approach
is due in large measure to the fact that one can speak, meaningfully, of
the probability of a theory, or of an hypothesis. 
 Moreover, since {\em Theory} can be anything
whatsoever one anticipates that 
the domain of applicability of Bayesian reasoning is 
considerable larger than that of a theory where the notion
of the probability of an hypothesis is absent, as is the case in
the frequentist approach.
However, this enormous conceptual gain comes at a
price. In order to arrive at
a posterior probability the price to be paid is
the specification of a prior probability for the
{\em Theory}, independently of the {\em Data}. There is simply no way
around this if one wishes to adhere to the rules of
probability theory. 

In many applications in high energy physics 
we are interested in propositions of the form
$\theta \in (a, b)$, that is, 
a parameter has a value within some continuous set.
Let
\seqn
\prob{\bx}{\theta, \lambda} = 
\int_{\Omega} \pdf{\bz}{\theta, \lambda} d\bz,
\eeqn
be the probability assigned to the data-set $\bx$, contained in a
neighborhood $\Omega$ of $\bx$, 
and let $\theta$ and $\lambda$ be the parameters
of the model currently under
consideration. Perhaps $\theta$ is the parameter
of interest, say the mass of the Higgs boson,
while $\lambda$ represents parameters such as the mean background
rate and the jet energy scale. It could even represent purely
theoretical parameters, such as the renormalization and 
factorization scales. All such parameters, which are not of intrinsic
interest, are referred to as {\bf nuisance parameters}.

If $\prior{\theta, \lambda} = \priorpdf{\theta, \lambda} d\theta d\lambda$ 
is the prior probability assigned to
the proposition that $\theta$ and $\lambda$ have certain values
 --- where $\priorpdf{\theta, \lambda}$ is the {\bf prior density}, we can
write Bayes' theorem as
\seqna
    \prob{\theta, \lambda}{\bx} & = & 
    \frac{  \prob{\bx}{\theta, \lambda} \, \prior{\theta, \lambda} }
         {  \int_{\theta, \lambda}
            \prob{\bx}{\theta, \lambda} \, \prior{\theta, \lambda} } \, ,
        \nonumber \\
        & = & \pdf{\theta,  \lambda}{\bx} \, d \theta \, ,
\eeqna
which in terms of densities becomes
\seqn
\pdf{\theta,  \lambda}{\bx} = 
    \frac{  \pdf{\bx}{\theta, \lambda} \, \priorpdf{\theta, \lambda} }
         {  \int d\theta \int d\lambda
            \pdf{\bx}{\theta, \lambda} \, \priorpdf{\theta, \lambda} }\, .
\eeqn
Since the nuisance parameters $\lambda$ are not of interest we need a
way to get rid of them in order to say something useful about the
parameter that is. 
This is technically difficult in the frequentist
approach, but straightforward in principle in the Bayesian approach: one
``merely'' integrates them out of the problem
\seqn
\label{eq:PosteriorDensity}
    \pdf{\theta}{\bx} = \int \pdf{\theta, \lambda}{\bx} \, d\lambda.
\eeqn
The quotation about the word merely is appropriate because it may
be difficult, in practice, to perform what are often high-dimensional
integrals. That being said, the posterior density, 
\Eq{PosteriorDensity}, is an
elegant encapsulation of all that
we know about the parameter $\theta$, given the data we have
acquired and the prior knowledge encoded in the prior density
$\priorpdf{\theta, \lambda}$.

\subsection{The Likelihood Principle}
\label{sec:LikelihoodPrinciple}
The posterior density, 
$\pdf{\theta}{\bx}$ --- the final result of
our inference about $\theta$,
displays a very important philosophical, and
practical, difference between the frequentist and Bayesian approaches that
we have alluded to, namely, that in a Bayesian analysis
\begin{quote}
an inference 
depends only on the data observed,
\end{quote}
a principle that is referred to as
the {\bf likelihood principle}, not to be confused with
the method of maximum likelihood. 
 Clearly, to base
an {\em inference} on an
ensemble of possible data-sets is to be
sharply at odds with the likelihood
principle. Consequently, the
principle is at odds with a
 host of standard frequentist practice. Since these
methods are still firmly entrenched, 
one is naturally led to ask:
is the likelihood
principle sensible? Certainly, this was Jeffreys~\cite{Jeffreys} opinion.
Ironically,  even Fisher --- a
forceful critic of all things Bayesian --- was an advocate of the
likelihood principle. Indeed, Fisher was extremely critical
of what he regarded as the ``extreme frequentism'' advocated 
by Neyman. 
A further irony is that, according to a theorem due to
Birnbaum~\cite{Birnbaum},
the likelihood principle follows from ideas that many frequentist
statisticians consider unimpeachable.

\subsection{Parameter Estimation}
\label{sec:ParameterEstimation}
The posterior probability is a complete statement of 
the results of an inference. However, particular summaries are often
of direct interest. Having finally arrived at a posterior
density for the Higgs boson mass, what we want, of course, is a
single mass estimate plus some idea of how well the mass has been
measured. In some circumstances,
it may
be useful to take the mean of the posterior
density as an estimate of
the parameter of interest. However, the mean is not the only
possibility.
One way to formalize the
construction of estimates is through loss functions, which we
discussed in general terms in \Sec{LossAndRisk} and which we
discuss in more detail below.

In the Bayesian approach it is natural to
speak of our {\em knowledge} being uncertain, in particular,
our knowledge of the value
of a parameter. Moreover, the uncertainy in our knowledge is
measured not by the expected
scatter of estimates over an ensemble, as would be the case
in a frequentist analysis, but rather by 
some measure of the width of the posterior density,
which, 
in accordance with
the likelihood principle, 
depends only on the observed data.
       
As noted above, a loss function is 
a way to measure the quality of a decision. 
A typical decision is:
given a data-set $\bx$
decide that the estimate of $\theta$ is $\htheta = d(\bx)$, where
$d(\bx)$ is a special kind of decision function called an estimator.
To illustrate these ideas, we consider two commonly
used loss functions. 

\subsubsection{Quadratic Loss}
\label{sec:QuadraticLoss}
The quadratic loss, introduced earlier, is
\seqn
    \loss{\theta,d} = \lr{\theta - d}^{2} \, . \label{eq:qloss}
\eeqn
Earlier, we also introduced the average loss, that is, the
risk function.
In the frequentist theory, the averaging is done with respect to
an ensemble of possible data-sets $\bx$.
In the Bayesian theory, one
averages over all
possible propositions about the value of $\theta$,
constrained by the fact that we have a obtained a specific
data-set. Therefore,
we are led to consider the risk function
\seqna
    \risk(\bx)  & = & \Ave{\loss{\theta,d}}{\theta}, \label{eq:msed} 
    \nonumber \\
                & = & \int \, \loss{\theta,d} \pdf{\theta}{\bx} d\theta,
\eeqna
that is, 
\seqn
    \risk(\bx)  = \int \, (\theta - d)^2 \pdf{\theta}{\bx} d\theta,
\eeqn
for the quadratic loss, where $\pdf{\theta}{\bx}$ is the posterior
density.  The best estimator
is declared to be that which minimizes the risk
\seqna
    \D{d} \risk(\bx)  & = &
    \D{d} \int \, \loss{\theta, d} \pdf{\theta}{\bx} d\theta,
    \nonumber \\
        & = & \int \, \D{d} \loss{\theta, d} \pdf{\theta}{\bx} d\theta,
    \nonumber \\
        & = & 0.
\eeqna
To simplify the notation, we use 
the symbol $\D{d}$ to represent the derivative with respect to $d$.
(Also, being physicists, we naturally assume that the derivative 
and integral operators commute.)
After minimization, we obtain
the intuitively pleasing result
\seqn
    \htheta = d(\bx) = \int \, \theta \, \pdf{\theta}{\bx} d\theta.
\eeqn
In words: 
\begin{quote}
The optimal estimate with respect to 
a quadratic loss is
the mean of the posterior density. 
\end{quote}

\subsubsection{Absolute Loss}
\label{sec:AbsoluteLoss}
The absolute loss, defined by
\seqn
    \loss{\theta,d} = |\theta - d| \, , \label{eq:absloss}
\eeqn
is used when one wishes to be more tolerant of 
deviations from the mean. Estimates based on the absolute loss
are less sensitive to the
tails of the posterior density and in that sense are more robust
than those based on the quadratic loss. As before,
we obtain the estimator $d$ by minimizing the risk 
\seqn
    \risk(\bx)  = \int \, |\theta - d| \, \pdf{\theta}{\bx} d\theta.
\eeqn
Differentiating with respect to the function $d$ yields
\seqna
    D_d \risk(\bx)  & = & 0 \nonumber \\
        & = & \int \, D_d |\theta - d| \, \pdf{\theta}{\bx} d\theta
        \nonumber \\
        & = & - \int \, 
        \frac{\theta-d}{|\theta - d|} \, \pdf{\theta}{\bx} d\theta \, ,
\eeqna
that is,
\seqn
        \int_{\theta < d} \, \pdf{\theta}{\bx} \, d \theta = 
        \int_{\theta > d} \, \pdf{\theta}{\bx} \, d \theta \, ,
\eeqn
which shows that the optimal estimator $d$, using the absolute loss,
 is the {\bf median} of the posterior density.

\subsubsection{Uncertainty}
\label{sec:Uncertainty}
The uncertainty in our knowledge of a parameter is quantified
by some measure of the
width of the posterior density.  
One such measure is
the variance 
\seqn
    \var{\theta} = \ave{\theta^2} - \ave{\theta}^2.
\eeqn
Another is a {\bf credible interval}, $\cin{\bx}$,
referred to also as a Bayesian interval,
obtained from the formulae
\seqn
    \int_{\theta \leq l(\bx)} \, \pdf{\theta}{\bx} d\theta = \alpha_{L}
\eeqn
and
\seqn
    \int_{\theta \geq u(\bx)} \, \pdf{\theta}{\bx} d\theta = \alpha_{R},
\eeqn
where
$\alpha_{L}$ and $\alpha_{R}$ as chosen so that 
$\beta = 1 - \alpha_{L} - \alpha_{R}$, where $\beta$ is the desired
probability, that is, degree of belief, to be assigned to the
specified interval.
The interpretation of credible intervals is direct:
$\beta$ is the probability that the proposition
$\theta \in \cin{\bx}$ is true.

\subsection{Combining Results}
\label{sec:CombiningBayesian}
In the frequentist approach the results from different experiments are
combined using a weighted average. However, more generally, results can be
combined using 
Bayes' theorem.
Let $\pdf{\bx_k}{\theta, \lambda, \alpha_k}$ be the likelihood for experiment
$k$, where $\theta$ is the parameter of interest and $\lambda$ represents
any nuisance parameters that are common to all experiments --- this could
be, for example, a measured cross section used by all experiments --- and
$\alpha_k$ represents nuisance parameters specific to experiment $k$. Ideally,
for each experiment the {\bf marginal likelihood},
\seqn
        \pdf{\bx}{\theta, \lambda} = \int 
        \pdf{\bx}{\theta, \lambda, \alpha_k} \, 
	\priorpdf{\alpha_k} \, d\alpha_k 
        \, ,
\label{eq:marginal}
\eeqn
 would be reported, that is, the likelihood function marginalized with
respect to the nuisance parameters 
$\alpha_k$ specific to the experiment. We do not marginalize, at this
stage, with
respect to $\lambda$ because these parameters are common across experiments.
The function $\priorpdf{\alpha_k}$ is
the prior density for $\alpha_k$.
In writing \Eq{marginal},
we have implicitly factorized the full prior density 
$\priorpdf{\theta, \lambda, \alpha_k}$ as follows
\seqn
\priorpdf{\theta, \lambda, \alpha_k}  =  
        \priorpdf{\theta, \lambda | \alpha_k} \, \priorpdf{\alpha_k} .
\eeqn
We shall assume that for every experiment, whose results
are to be combined, the prior density 
$\priorpdf{\theta, \lambda | \alpha_k}$ is 
independent of $\alpha_k$, in which case
we may write
 \seqn
\priorpdf{\theta, \lambda, \alpha_k}  =  
        \priorpdf{\theta, \lambda} \, \priorpdf{\alpha_k}.
\eeqn
Given this assumption, each experimental group, if it wishes, can produce 
an inference about $\theta$ and $\lambda$ by 
supplying a prior density $\priorpdf{\theta, \lambda}$. This observation
provides the clue
about how to combine results. The prior density 
$\priorpdf{\theta, \lambda}$ for
a given experiment is simply the posterior density 
$\pdf{\theta, \lambda}{\bx}$ from another.  
Therefore, by recursively combining
the results from $K$ experiments
we obtain the overall  posterior
density
\seqn
    \pdf{\theta, \lambda}{\bx_1,\ldots,\bx_K} = 
    \frac{  \pdf{\bx_1}{\theta, \lambda} \cdots
      \pdf{\bx_K}{\theta, \lambda} \priorpdf{\theta, \lambda}
    }
         {  
           \int d\theta \int d\lambda \, \pdf{\bx_1}{\theta, \lambda} \cdots
           \pdf{\bx_K}{\theta, \lambda} \priorpdf{\theta, \lambda} }.
\eeqn
This is proportional to the product of the joint
likelihood function for the combined results  and a 
prior density for $\theta$ and $\lambda$.
This method will yield estimates that
converge to the true value as more and more experiments are combined,
provided that the result 
from each experiment is {\bf consistent}. By consistent we
mean that the estimates from an experiment would converge to the
true value, as more and more data are acquired in that experiment, 
with a probability that
approaches unity. Note that a consistent estimator need not be unbiased.
However, by definition, its bias vanishes in
the limit of large data-sets.

\subsection{Model Selection}
\label{sec:ModelSelection}
Suppose we have a set of competing models $M$,
which may depend upon different sets
of parameters
$\theta_M$ and we wish to pick the one that fits the data best. 
Given some prior information and a
data-set $\bx$, how should one make this decision?
This is the problem of {\bf hypothesis testing} or {\bf model selection}. 

Our first task 
is to assign a probability density, $\pdf{\bx}{\theta_M, M}$,
 to our data-set given a model $M$ and
hypotheses about the values of the corresponding
parameters $\theta_M$. We must
also assign a prior density $\priorpdf{\theta_M, M}$. Then
write down Bayes' theorem
\seqn
    \pdf{\theta_M, M}{\bx} = 
    \frac{\pdf{\bx}{\theta_M, M} \, \priorpdf{\theta_M, M}}
     {\sum_M \, \int \,
          \pdf{\bx}{\theta_M, M} \, \priorpdf{\theta_M, M} \, d\theta_M}.
\eeqn
The function $\pdf{\theta_M, M}{\bx}$ represents the
probability density of the proposition: $M$ is the true model and it
has parameter values $\theta_M$.

It is very important to understand 
that the probability densities $\pdf{\theta_M, M}{\bx}$ 
are conditioned on
the set of models considered, {\em so far}.
``Best model'' in this context simply means the best of the current set.
Should another model be added to the set,
the probabilities assigned to different models 
would, in general, change. Therefore, 
$\pdf{\theta_M, M}{\bx}$ cannot be
construed as an absolute measure of the validity of a 
model. But it {\em is} a measure of
the {\em conditional} validity of a model:
it provides a way to {\em compare} models within a 
given set in light of what we know. If a rational thinker
had to choose a single model she 
would opt for the model with the highest
posterior probability. But, should she acquire further pertinent
information, that information, via Bayes' theorem, could cause her
to change her mind about which model is currently best. 

Finally, we can marginalize $\pdf{\theta_M, M}{\bx}$
with respect to $\theta_M$ to obtain $\prob{M}{\bx}$,
the probability of model $M$. This is
potentially very useful if each model, within the set, are
identical, except for the value of a single parameter $\alpha$. 
For example, $M$ could
label models that differ by an assumed value for the mass of the
Higgs boson.
We then have a way to estimate that parameter:
\seqn
    \hat{\alpha} = \sum_M \alpha_M \prob{M}{\bx},
\eeqn
and its associated uncertainty
\seqn
    \sigma_{\alpha}^2 = \sum_M (\alpha_M-\hat{\alpha})^2 \prob{M}{\bx}.
\eeqn

\subsection{Optimal Event Selection}
Before we can measure something, we must find a it.
Therefore, a basic task of data analysis is 
to separate signal from background.
Given a set of discriminating variables, 
the traditional method combines
a judicious  use of common sense, physical intuition, and 
trial and error to separate signal from background. 
However, much of the energy devoted to this can be
better spent elsewhere since the task of finding the
optimal separation between signal and background is a
well-defined mathematical problem whose solution is known.

It helps to think about the problem geometrically. Suppose we have
found $n$ variables that we consider useful for separating signal
from background. The $n$ variables can be thought of as a point in an
$n$-dimensional space, sometimes referred to as {\bf feature space}.
Presumably, by construction, the signal tends to cluster in one part
of this space while the background tends to occupy a different region. 
However, inevitably, there
will be some overlap between the signal and background densities. The
problem to be solved is to find the boundary that separates
optimally 
signal from background. Tradionally, one does 
the simplest thing: one constructs a
boundary from planes that are perpendicular to the axes, where each plane
corresponds to a cut on a specific variable.  
However, in general, the optimal boundary cannot be built from
such intersecting planes; in general, it
will be a curved surface. 
 
The problem of finding this surface, however, is indeterminate until we
have specified what
we mean by optimal. A generally accepted definition of an optimal
boundary is
one
that minimizes the probability to
misclassify events.
For the moment,
we shall suppose
that we know the signal and background densities, $f(\bx|S)$ and
$f(\bx|B)$, respectively. Let us further assume that we know the signal
and background prior probabilities $P(S)$ and $P(B)$. These prior
probabilities are not controversial: $P(S)$ is just the chance to pick
a signal event without regard to its feature vector
$\bx$, and likewise for $P(B)$. Since the event must be either signal or
background it must be the case that
$P(S) + P(B) = 1$. The probability to misclassify 
a signal event, with feature vector $\bx$, is just
the probability for signal events to land on the
background side of the optimal boundary, or for a background
event to land in the signal region. For simplicity, 
we consider a one dimensional problem, with the
boundary, say, at $x = x_0$. The probability $E_S$ to misclassify
a signal event is
\seqn
    E_S(x_0) = P(S) \int h(x_0 - x) f(x|S) dx,
\eeqn 
where $h(z)$ is the Heaviside step function, 
defined by $h(z) = 1$ if $z > 0$ and
zero otherwise. The probability to misclassify the background is, likewise,
the probability for the background to land on the signal side,
\seqn
    E_B(x_0) = P(B) \int h(x - x_0) f(x|B) dx.
\eeqn
Hence, the probability to misclassify events, that is, the {\bf error rate},
regardless of whether they are signal or background, is the sum
\seqn
E(x_0) = E_S(x_0) + r \, E_B(x_0),
\eeqn
where $r$ is a weight that allows
 for the possibility that we may wish to weight the
background more (or less) than the signal. We now minimize $E(x_0)$ with
respect to the choice of boundary, that is, we set $\D{x_0} = 0$ and
obtain 
\seqn
p(S) \int \delta(x_0 - x) f(x|S) dx + P(B) \int \delta(x - x_0) f(x|B) dx = 0.
\eeqn
The derivative has conveniently converted the step functions into delta
functions, thereby rendering the integrals trivial, yielding
the result
\seqn
        r(x_0) = \frac{f(x_0|S) P(S)}{f(x_0|B) P(B)}.
\eeqn
The function $r(*)$ is called the {\bf Bayes discriminant} because of
its intimate connection with Bayes' theorem,
\seqn
P(S|x) = \frac{r}{1+r} = \frac{f(x|S) P(S)}{f(x|S) P(S) + f(x|B) P(B)}.
\label{eq:psx}
\eeqn
The $n$-dimensional generalization of this has the same Bayesian form.
(See Ref.~\cite{Barlow2} for an interesting derivation of this result.)
The posterior probability $p(S|\bx)$ is precisely that needed for event
classification. It is the probability that an event characterized by the
vector $\bx$ is of the signal class. By using this
probability we have succeeded in mapping the original $n$-dimensional
problem into a more tractable one-dimensional one.

This is all very well, but there is a
serious practical problem. Rarely do we have analytical expressions 
for the signal and background densities $f(\bx|S)$ and $f(\bx|B)$.
We seem, alas, to have achieved a pyrrhic victory! Happily,
however, many methods exist that provide good approximations to the
posterior probability. In particular, it has been
shown that, under suitable circumstances, neural networks~\cite{nn}
compute a direct approximation to the probability $p(S|\bx)$.
 
\subsection{Prior Probabilities}
So far, we have skirted over a potentially serious difficulty of
the Bayesian approach;
to solve an inference problem we must assign two quantities, 
a prior and a likelihood.
There is broad agreement within physical sciences about the
use of a Poisson distributions to model counting experiments.
However, 
even amongst those who agree
that prior probabilities are necessary, there is disagreement
about how to
assign them when we have
minimal prior information about the parameters
to be estimated, or when we wish to act as if this were so. 
The basic problem is to assign a prior that, in some well-defined sense,
has as small an effect as possible on the final inference. In other
words, most physicists want a method that ``let's the data speak
for themselves''. At face value, this is the strength of
the frequentist approach where no priors appear. However, this
strength is illusory because it forces one to answer the wrong question,
namely, given a particular model $M$ one is forced to answer the
question: what data-sets are possible? But, the
question of direct interest is the inverse: given a particular data-set,
namely, the one actually obtained, what models are compatible with it?

A Bayesian analyst is often faced with the following
circumstance: that the only prior information at hand about a
parameter $\theta$ is that it lies within some set, perhaps the set
$\theta \in [0,\infty)$.
What prior probability should we assign to various hypotheses about
its value? Laplace argued that if we
know nothing about the value of a parameter then we should assign a
flat prior density to encapsulate this state of knowledge: 
$\priorpdf{\theta} \propto \, \mbox{constant}$.
This seems reasonable, until we realize that {\em any} choice of
prior density for a given parameter $\theta$ specifies, implicitly,
the prior density for the infinity of
parameters that are functions of $\theta$. Clearly, we have
specified a lot more than we bargained for!

For example, suppose we transform from $\theta$ to the
parameter $\alpha = 1/\theta$. Inferential coherence
demands that its prior probability density be 
$\priorpdf{\alpha} \propto 1 / \alpha^2$;
a form that looks, at best, non-intuitive.
This prior density would be fine were it not for
the following question: what reason do we
have to suppose that the prior density is flat in the parameter
$\theta$ rather than in the parameter $\alpha$, or
some other parameter,  such as $\tau = \ln \theta$? It seems that the
assignment of prior probabilities for a parameter about which we are
almost totally ignorant is, indeed, arbitrary. This in
a nutshell is the core of the 
controversy about prior probabilities that has raged
for more than 200 years.

The problem of how to assign prior probabilities that, in some
sense, have the smallest effect on inferences has a long, 
difficult, and polemic history~\cite{Wasserman}. Here, however, is
some practical advice. Use the prior density
that seems most reasonable to you or, 
better still, one that has been agreed upon by
the community for the given problem. For example, both
the CDF and D\O\ Collaborations have agreed to use a 
flat prior for a cross-section. Then check the robustness of the
inferences (that is, see how much they vary) 
by trying different reasonable priors. If the answers are
unduly sensitive  to the choice of prior  then
the scientifically honest conclusion should be that the data at hand are
inadequate and more should be acquired.

\subsection{Counting Experiments}
\label{sec:BayesianAnalysis}
We have covered the basic elements of the Bayesian theory. In
this section, we illustrate some of this theory by applying it to 
a prototypical example in high energy physics: the analysis of
a counting experiment.

Every Bayesian analysis contains at least four ingredients:
\begin{itemize}
\item A model
\item A data-set
\item A likelihood
\item A prior probability
\end{itemize}
For a counting experiment the model is
\seqn
a = s + b,
\eeqn
where $a$ is the mean number of events, $s$ the mean signal count
and $b$ the mean background count. Let $n$ be the total number of events
observed. As discussed in Lecture 2, the probability to observe $n$ events
may be assumed to be 
\seqn
\prob{n}{s, b} = \Poisson{n}{s+b}.
\eeqn
The prior density for $s$ and $b$ can be factorized thus
\seqna
\priorpdf{s, b} & = & \priorpdf{s|b} \, \priorpdf{b},
\nonumber \\
& = & \priorpdf{s} \, \priorpdf{b},
\eeqna
where we have assumed that the conditional prior density for the signal
does not depend on the value of the background. We have two prior
densities to assign. We consider first the prior for the background, then
that for the signal.

Let us suppose that the background has been estimated from a Monte
Carlo simulation of the background process, yielding $B$ background
events, with probability given by
$\prob{B}{\lambda} = \Poisson{B}{\lambda}$. 
Furthermore, we assume that the relationship
between $b$ and $\lambda$ is 
\seqn
b = k \lambda,
\eeqn
where $k$ is a known scale factor, in this example, the ratio 
of the observed to Monte Carlo integrated luminosities. Given
$B$, we can compute the posterior density,  
\seqn
    \pdf{\lambda}{B} = 
    \frac{  \pdf{B}{\lambda} \, \priorpdf{\lambda} }
         {  \int \pdf{B}{\lambda} \, \priorpdf{\lambda} \, d\lambda } ,
\eeqn
for $\lambda$. But, to do so requires specification of the prior
density $\priorpdf{\lambda}$. We shall suppose that
it is of the form $\priorpdf{\lambda} = \lambda^p$, but, for simplicity, we
consider $p = 0$, that is, a flat prior in $\lambda$. The posterior density
$\pdf{\lambda}{B}$ contains information about the parameter $b$,
by virtue of the relation $b = k \lambda$. It can therefore
serve as the {\em prior density} for $b$.
From Bayes' theorem we obtain the posterior density
\seqn
    \pdf{s, k \lambda}{n} = 
    \frac{ \pdf{n}{s, k \lambda} \, \priorpdf{k \lambda} \, \priorpdf{s} }
         {  \int \int 
	   \pdf{n}{s, k \lambda} \, \priorpdf{k \lambda} \, \priorpdf{s}
	     d\lambda \, ds},
\eeqn
from which we can
eliminate the nuisance parameter $\lambda$ by marginalization
\seqn
    \pdf{s}{n} = \int \pdf{s, k \lambda}{n} d\lambda.
\eeqn
The function $\pdf{s}{n}$ suggests that it may be convenient to define
the {\bf marginal likelihood}
\seqn
\pdf{n}{s} \equiv \int \pdf{n}{s, k \lambda} \, \priorpdf{k \lambda} \, 
d\lambda , 
\eeqn
and write Bayes' theorem as
\seqn
    \pdf{s}{n} = 
    \frac{ \pdf{n}{s} \, \priorpdf{s} }
         {  \int \pdf{n}{s} \, \priorpdf{s} \, ds }.
\eeqn
For this problem, the marginal likelihood can be calculated. The
result is
\seqn
\pdf{n}{s} =  \frac{1}{(1+k)^{B+1}}
\sum_{r=0}^n  \lr{\frac{k}{1+k}}^{n-r} \frac{\Gamma(n-r+B+1)}{(n-r)! B!} \,
\Poisson{r}{s}.
\eeqn

We now turn to the signal prior $\priorpdf{s}$. Our knowlege of the
signal is rather vague: we know it is positive and finite! It is far
from clear how to translate this prior knowledge into a prior density.
We shall simply adopt as a matter of {\em convention} the prior
$\priorpdf{s} = 1$. In practice, one gets intuitively reasonable 
results with it; 
but there are better choices~\cite{Wasserman}. Putting all
pieces together we can compute the posterior density $\pdf{s}{n}$, which
is the final, and complete, encoding of our improved knowledge of the
possible values of the mean signal count $s$.
\vskip 24pt
\framebox{{Exercise}: Derive the formulae for $\pdf{n}{s}$ and $\pdf{s}{n}$.}

\newpage
\section{Lecture 4 - Statistical Infserence, Part II}

\subsection{Goodness of Fit}
Consider the task of fitting a curve
to a histogram of counts. The usual way to do this is by the method
of maximum likelihood. Let $f(x, \theta)$ be the curve to be fit by
adjusting the parameters $\theta$. We minimize the sum
\seqn
\sum_i \ln \Poisson{k_i}{f(x_i, \theta)} ,
\eeqn
which is equivalent to maximizing the joint likelihood of the counts,
with respect the parameters. Having found the best fit parameters,
it is considered sound practice to test the {\bf goodness-of-fit}. The
concept of goodness-of-fit was introduced by Fisher. The basic idea
is simple: one invents a measure of discrepancy $D(x)$ between the
fitted curve and the data such that large values of $D$ would tend to
cast doubt on the hypothesis
that the curve fits the data. One calculates the probability
density $f(D)$ of the discrepancy $D(x)$, in principle by the method
described in Lecture 2 but in practice by Monte Carlo simulation, and
one computes
\seqn
p = \int_{D > D_0} f(D) \, dD ,
\eeqn
the {\bf p-value} for the observed discrepancy $D_0$. Should that number
be judged too small, the fit is {\em rejected} as a {\em bad}
fit because the discrepancy is correspondingly too large. If on the
other hand $p$ is large, the fit may, or may not, be good! Suppose,
for example, that the
discrepancy is defined by the quadratic form
\seqn
D(x) = \sum_i (k_i - f(x_i, \theta))^2 / \sigma_i^2 ,
\eeqn
and we find $D(x) = 0$, and therefore $p=1$! This does not
necessarily imply a good fit; goodness-of-fit is a misnomer. These
tests should really be called ``badness-of-fit'' tests!

\subsection{Confidence Intervals}
The purpose of this section is to explain as clearly as possible
the important frequentist concept of a confidence interval.
Consider the following questions
\begin{itemize}
\item What is the mass of the $\tau$ neutrino?
\item What is the mass of the top quark?
\item What is the mass of the Higgs boson?
\end{itemize}
and the following tentative answers
\begin{eqnarray}
\label{eq:answers}
m_{\nu} &   < & 18.2  \, \mbox{MeV}, \nonumber \\
m_t &       = & 175.0 \pm 3.1 \, \mbox{GeV}, \nonumber \\
m_H &       > & 114.3 \, \mbox{GeV}.
\end{eqnarray}
The statements in \Eq{answers} are unsatisfactory because 
they fail to indicate how much {\em confidence} we
should place in them. In the absence of a convention, $m_t =
175.0 \pm 3.1$ GeV conveys no more information than does $m_t = 100 \pm
20$ GeV. The statements
\begin{eqnarray}
\label{eq:betteranswers}
m_{\nu} &   < & 18.2  \, \mbox{MeV, with CL = 0.950}, \nonumber \\
m_t &       = & 175.0 \pm 3.1 \, \mbox{ GeV, with CL = 0.683}, \nonumber \\
m_H &       > & 114.3 \, \mbox{GeV, with CL = 0.950},
\end{eqnarray}
are better because they assign probabilities,
called {\bf confidence levels} (CL) 
that indicate how seriously the statements should be
taken. If the statements, \Eq{betteranswers}, were Bayesian there is
nothing more to be said. The probabilities would be measures
of degrees of belief. However, since we wish to interpret them in
a frequentist manner, this involves a bit more work.

But first we ask the following question:
do the three statements in \Eq{betteranswers} convey
information that is different in kind? 
As written, the statements look
rather different. However, each can be re-expressed as follows
\begin{eqnarray}
m_{\nu} & \in & [0, 18.2] \mbox{ MeV, with CL = 0.950,} \\
m_t & \in & [169.2, 179.4] \mbox{ GeV, with CL = 0.683,} \\
m_H & \in & [114.3,\infty) \mbox{ GeV, with CL = 0.950,}
\end{eqnarray}
that is, as statements about intervals. 
Written this way,  it is clear that each
statement is conveying the same kind of information,
which loosely
speaking is this: a parameter of interest has a true {\em fixed}
value that has a good chance of being within the specified interval.
The second statement in \Eq{betteranswers} is an example of the
conventional way to state the results of a measurement. The number
$3.1$, associated with the confidence level of 0.683 
(or equivalently, 68.3\%), 
is referred to as a
{\bf standard error}. Sometimes (this is especially true in
searches for new phenomena) one is interested in only one
of the bounds of the interval. For example, in the first
statement in \Eq{betteranswers} the upper bound, that is, the {\bf upper
limit}, is of greater interest than the lower one.
In the third statement it is the lower bound that is of interest,
that is, the {\bf lower limit}.

\subsubsection{Coverage Probability}
Imagine a {\em set of ensembles} of experiments, 
each element of which is associated with a single {\em fixed} value $\theta$ 
of the parameter to be measured.
We might visualize each ensemble, within the set, as a huge box
filled with experiments, 
with each box labeled by (that is to say, associated with) a different
value of $\theta$. Each experiment $E$
yields an interval $[l(E), u(E)]$ for the parameter $\theta$. 
In each ensemble (that is, box) 
some fraction of the experiments will yield intervals that 
contain the $\theta$ value
associated with that ensemble. This fraction is called the {\bf
coverage probability}, which in general will vary from one
ensemble to another. 
The {\bf confidence level} is the {\em minimum
coverage probability} over the set of ensembles. In terms of our
fanciful picture, each box of experiments will have some coverage probability;
obviously, at least one box will have the smallest coverage probability,
which, by definition, is the confidence level of the set of boxes. 
Now suppose we
choose a box and repeatedly, and randomly, pick an experiment from it. We
shall find that the fraction of sampled experiments that yield
intervals containing $\theta$ is greater than or equal to the 
confidence level for our set of boxes.

The thought experiment we have just described cannot, of course, 
be carried out
in practice. However, coverage (as Neyman stressed) 
pertains not only to ensembles of
identical experiments measuring the same thing, but also to ensembles
of {\em different} experiments measuring {\em different} things. 
If one considers the (finite) 
ensemble of published intervals there is no doubt that
they have {\em some} coverage probability. But it is not clear
how useful it is to know this since we are not privy
to the true values of all the different quantities to which they pertain.
And if we were, the experiments would never have been undertaken!
The question remains, in 
what sense is a confidence level a measure of confidence, as commonly
understood? The basic idea is this: an experiment is imagined
selected at random from the ensemble (the box) to which it
belongs, presumably the one labeled by a $\theta$ whose value is equal
to that
dictated by  Nature. The probability that our experiment
yields an interval $[l(E), u(E)]$ containing $\theta$ is exactly
equal to the coverage probability of the ensemble to which it belongs, 
which by construction
is greater than or equal to the confidence level of the set of ensembles.
A confidence level is a measure of confidence in the sense that
the higher the confidence
level the more confident {\em we} are invited to be that {\em our} interval
actually contains the true value of $\theta$.

 We now can state the central problem
that must be solved in order to make probabilistic statements such
as those in \Eq{betteranswers} within the context
of a relative frequency interpretation of probability. 
The problem is to construct
intervals that {\em a priori} have a 
coverage probability equal to the desired confidence
level, or greater, 
{\em whatever the true value of the parameter of interest}. The
qualifying clause is necessary because we do not know what the true value is.
We do not know from
which box our experiment has come!
If a
set of intervals satisfies the above criterion they are said to {\bf
cover.} Intervals so constructed are called {\bf confidence
intervals}, a concept introduced by Jerzy Neyman 
in a seminal paper published in 1937~\cite{Neyman}. Actually, Neyman went 
further: he
required not only that confidence intervals cover for all possible
values of the parameter of interest but also for all possible values of
{\em all} the other parameters of the problem, commonly referred
to as  nuisance parameters. Again, this is necessary because we generally 
do not know their true values.

\subsubsection{The Neyman Construction}
In this section, we give the general algorithm for constructing
confidence intervals, which Neyman described in his 1937 paper. For
concreteness, we consider the problem of constructing
confidence intervals for the Poisson distribution
with mean count $\theta$.

\begin{figure}[htbp]
\includegraphics[width=\width]{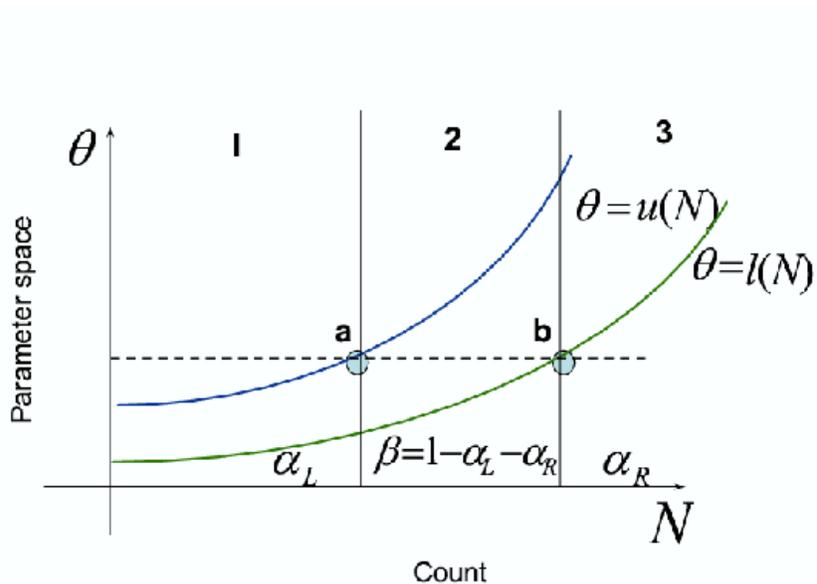}
\protect\caption{The plot shows how an observed count $N$ is
mapped into an interval $[l(N),u(N)]$, drawn vertically, 
in the parameter space of
$\theta$. As the count $N$ varies, so do the intervals. Counts
that land in region 2 lead to intervals that bracket the
true value of $\theta$, while counts that land in either region 1
or region 3 exclude $\theta$. But since the true value of $\theta$
is unknown we must construct the curves $\theta = u(N)$ and
$\theta = l(N)$ so that for every value of $\theta$ that is
possible {\em a priori}, that
is, for every box of experiments, the probability to
get a count in region 2 is $\geq \beta$. 
The points $a$ and $b$, respectively,
define the lower and upper bounds of an
interval in $N$ with probability content $\geq \beta$. }
\label{fg:cl}
\end{figure}

Happily, the algorithm is conceptually simple. It is illustrated in
Fig.~\ref{fg:cl}, which shows a plot of the parameter $\theta$
versus the observed count $N$.
Each point of the parameter space of $\theta$ is associated with
an ensemble of experiments, each yielding a count $N$ and an
interval $[l(N), u(N)]$, drawn vertically. The algorithm to construct
confidence intervals, when the probability density of
the observations depends on $\theta$ 
only  proceeds as follows. For
each value of $\theta$ one finds two counts $a$ and $b$, as indicated 
in Fig.~\ref{fg:cl}, such that the probability
to observe a count within the set $\{a,\ldots,b\}$ is $\geq
\beta$, where $\beta$ is the desired confidence level.
Figure~\ref{fg:cl} shows that, for a given $\theta$, the counts
$a$ and $b$ partition the space of observations into
three regions denoted 1, 2 and 3. If an experiment, from the
ensemble indexed by $\theta$, yields a count $N$
that lands in region 2 then the interval $[l(N), u(N)]$ will
bracket $\theta$. On the other hand, for all observations that fall in
either regions 1 or 3 the intervals will fail to include $\theta$.
By construction, the relative frequency with which a count falls in
region 2 is $\geq \beta$; therefore, the coverage probability of the
confidence intervals $[l(N), u(N)]$ will be {\em exactly} equal to
the probability to obtain a count in that region.

\subsubsection{Other Constructions}
There are many ways to
construct sets of counts $N$ with probability content
greater than or equal to the
desired confidence level simply by sliding the
points $a$ and $b$ along the horizontal line $\theta = \mbox{constant}$ 
(see Fig.~\ref{fg:cl}). 
One common way is to assign
equal probabilities $\alpha_L$ and $\alpha_R$ to the
regions 1 and 3, respectively.  Confidence intervals constructed
this way are called {\bf central confidence
intervals} and are most efficiently computed by solving the
equations 
\begin{eqnarray}
\label{eq:central}
    \alpha_L & = & \mbox{Pr}(r \leq N| \theta=u), \nonumber \\
            & = & \sum_{r=0}^N \Poisson{r}{u}, \nonumber \\ \\
    \alpha_R & = & \mbox{Pr}(r \geq N| \theta=l), \nonumber \\
             & = & \sum_{r=N}^{\infty} \Poisson{r}{l}, \nonumber \\
             & = & 1 - \sum_{r=0}^{N-1} \Poisson{r}{l},
\end{eqnarray}
where $\beta = 1 - \alpha_L - \alpha_R$, with $\alpha_L$ set equal
to $\alpha_R$. (The subscript $L$ stands for left and $R$ for right,
corresponding to the regions left and right of region 2 in Fig.~\ref{fg:cl},
that is, regions 1 and 3, respectively.)

Another method that has gained adherents is that of
Feldman and Cousins~\cite{FeldmanCousins}. In this method, as in the
general case, one finds for each value of $\theta$ a set of counts
$\{N\}$ such that the probability to obtain a count within the set is
$\geq \beta$. The set is populated by first ordering $N$ according to
the likelihood ratio
\begin{equation}
    \frac{\Poisson{N}{\theta}}{\Poisson{N}{N}},
\end{equation}
in descending order, and then adding values of $N$ to the set until its
probability content is equal to or just exceeds the desired
confidence level. The counts $a$ and $b$ are the
minimum and maximum values within the set $\{N\}$. A procedure for
populating sets of observations, such as $\{N\}$, with specified
probability content is called an {\bf ordering principle}. 
The one just described is referred to as Feldman-Cousins ordering.

\begin{figure}[htbp]
\includegraphics[width=\width]{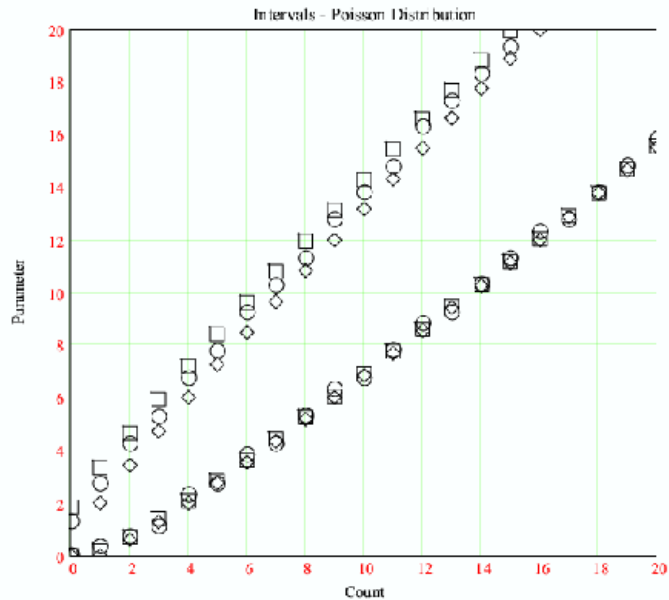}
\protect\caption{Confidence intervals for the Poisson
distribution. Three sets of intervals are shown: central intervals
(boxes), Feldman-Cousins intervals (circles) and ``root N''
intervals (diamonds).} \label{fg:intervals}
\end{figure}

\begin{figure}[htbp]
\includegraphics[width=\width]{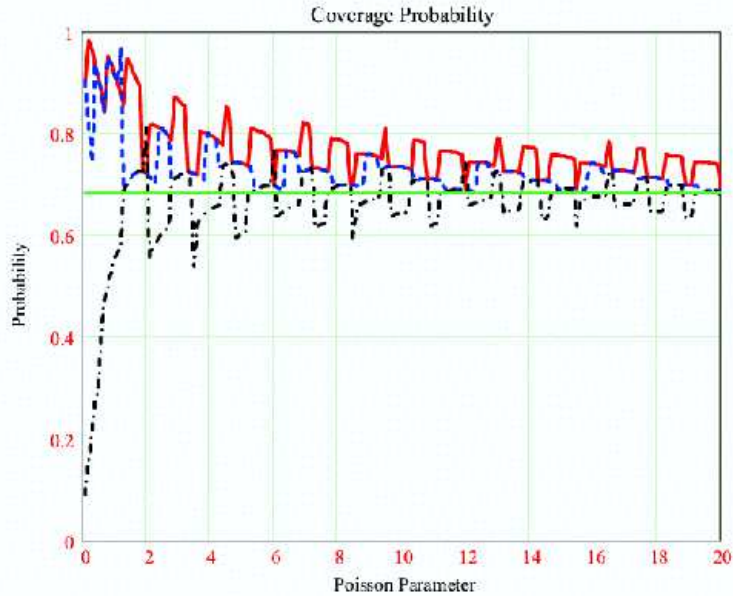}
\protect\caption{Coverage probability for three sets of Poisson
confidence intervals, central (solid line), Feldman-Cousins
(dashed line) and ``root N'' (dot-dashed line), as a function of
the parameter $\theta$. By construction the central
and Feldman-Cousins intervals cover, whereas the simple ``root N''
ones do not.} \label{fg:coverage}
\end{figure}

Figure~\ref{fg:intervals}
compares central intervals with those constructed using the
Feldman-Cousins method. We also show the
intervals given by the well-known ``root N'' rule $l(N) = N-\sqrt N$ and
$u(N) = N+\sqrt N$. We see that all three intervals have
approximately the same lower confidence limits, but that the
upper limits of central intervals are higher than those of 
Feldman and Cousins, which in turn are higher than those of the ``root N''
intervals. However, while both the central and Feldman-Cousins
intervals cover, as they necessarily must in view of how they are
constructed, the simple ``root N'' intervals do not, as indicated
in Fig.~\ref{fg:coverage}.
Note, however, that as $N \rightarrow \infty$ 
the ``root N'' intervals become 
ever more satisfactory approximations to the exact intervals. Incidentally,
the use of a confidence level of 0.683 stems from the fact that for 
$x \sim  \mbox{Gaussian}(x,\mu,\sigma)$, with mean $\mu$ and 
standard deviation $\sigma$, 
intervals of the form $[x - \sigma, x + \sigma]$ have
a confidence level of 0.683. The ``root N'' intervals converge to the Gaussian
ones as $N \rightarrow \infty$. 

Clearly there is considerable freedom of
choice in constructing confidence intervals. Consequently, with
exactly the same data different physicists within a collaboration
could compute different confidence intervals all of which cover.
So how is one to decide which interval to publish? Unfortunately,
there is no consensus, as yet, on the criteria to be used to select a
set of confidence intervals from the (infinite) set of possibilities.
The only non-controversial advice that can be given is this: in
a publication explain precisely what you have done!


\section*{Acknowledgements}
I wish to thank Prof.~Suman Beri for hosting such a memorable,
and enjoyable, school as well as all the students who, with their
youthful enthusiasm, made it
so worthwhile.


\end{document}